# Type I Error Rates are Not Usually Inflated

*Mark Rubin*
*Durham University, UK*

**Abstract**

The inflation of Type I error rates is thought to be one of the causes of the replication crisis. Questionable research practices such as *p*-hacking are thought to inflate Type I error rates above their nominal level, leading to unexpectedly high levels of false positives in the literature and, consequently, unexpectedly low replication rates. In this article, I offer an alternative view. I argue that questionable and other research practices do not usually inflate relevant Type I error rates. I begin by introducing the concept of Type I error rates and distinguishing between statistical errors and theoretical errors. I then illustrate my argument with respect to model misspecification, multiple testing, selective inference, forking paths, exploratory analyses, *p*-hacking, optional stopping, double dipping, and HARKing. In each case, I demonstrate that relevant Type I error rates are not usually inflated above their nominal level, and in the rare cases that they are, the inflation is easily identified and resolved. I conclude that the replication crisis may be explained, at least in part, by researchers' misinterpretation of statistical errors and their underestimation of theoretical errors.

During significance testing, a Type I error occurs when a researcher decides to reject a true null hypothesis (Neyman & Pearson, 1928, p. 177; Neyman & Pearson, 1933, p. 296). Type I errors are thought to play an important role in explaining the replication crisis in science. When the results of a study fail to replicate, the replication failure may be attributed to a Type I error in the original study. In other words, one of several reasons for a failed replication is that the null hypothesis is true and the original study's significant result was a false positive (Nosek et al., 2022, p. 726).

The replication "crisis" occurred because replication rates were lower than "expected or desired" (Nosek et al., 2022, p. 724; see also Munafò et al., 2017, p. 1; Open Science Collaboration, 2015, p. 7). Unexpectedly low replication rates have been attributed to larger than expected Type I error rates which, in turn, have been attributed to the use of questionable research practices. For example, Simmons et al. (2011) argued that "despite empirical psychologists' nominal endorsement of a low rate of false-positive findings (≤ .050), flexibility in data collection, analysis, and reporting dramatically increases actual false-positive rates" (p. 1359). Hence, questionable research practices are thought to inflate actual Type I error rates above the nominal conventional level of .050, leading to an unexpectedly high level of false positives in the literature and, consequently, unexpectedly low replication rates.

In this article, I offer the alternative view that relevant Type I error rates are not usually inflated by questionable research practices (e.g., *p*-hacking) or other research practices (e.g., model misspecification, exploratory analyses). I agree that Type I errors can be responsible for some replication failures. However, I argue that Type I error rate *inflation* is relatively rare, and that when it does occur it is easily identified and resolved. I conclude that theoretical errors provide a better explanation of the replication crisis than Type I error rate inflation.

My argument is consistent with philosophies of science and statistics that consider the logical relations between a hypothesis and its test result in the context of justification independent from their psychological origins, including researcher biases in the context of discovery (Popper, 1962, p. 140; Popper, 2002 p. 7; Reichenbach, 1938, p. 5). Based on this "logico-historical" view (Musgrave, 1974, p. 3), I argue that "actual" Type I error rates can be logically deduced from reported statistical inferences without referring to planned testing procedures that include the way in which hypotheses and evidence are constructed and selected. Consequently, my logical inference-based approach diverges from Mayo's (1996, 2018) error statistical approach, and it does not require preregistration to assess Type I error rates (Rubin, 2020a).

I begin with an introduction to Type I error rates that distinguishes them from theoretical errors. I then consider a range of questionable and other research practices that are thought to inflate Type I error rates. In particular, I consider model misspecification, multiple testing, selective inference, forking paths, exploratory analyses, *p*-hacking, optional stopping, double dipping, and HARKing. I demonstrate that relevant Type I error rates are not usually inflated by these practices. I conclude by summarizing my arguments, discussing the evidence for Type I error rate inflation, and considering some implications for our understanding of the replication crisis.

Before proceeding further, it is worth addressing a few potential misconceptions about my position. I am not arguing that questionable research practices do not exist. There is clear evidence that researchers sometimes engage in questionable research practices (e.g., John et al., 2012). My point is that these practices do not usually inflate Type I error rates. I am also not arguing that researchers are always honest or that they do not exploit researcher degrees of freedom. For example, I accept that researchers may falsely portray a post hoc hypothesis as an a priori hypothesis. However, I argue that this and other questionable research practices do not usually inflate Type I error rates. Furthermore, none of the points that I make condone dishonesty and the nondisclosure of potentially relevant information. Again, they only support the argument that questionable and other research practices do not usually inflate relevant Type I error rates. Finally, I am not denying that some of these research practices can be problematic. However, my argument is that they may increase *theoretical errors,* rather than *statistical errors*. In other words, they may lead to theoretical misinterpretation rather than Type I error rate inflation.

# Introduction to Type I Error Rates

## What Is a Type I Error Rate?

A Type I error rate is the frequency with which a researcher would decide to reject a true null hypothesis when they base their decision on the results of a significance test that is performed on a long run of random samples that are drawn from a null population in an imaginary situation in which random sampling error is the only source of error. There are four points to note about this definition.

First, the word “population” refers to not only a population of research participants, but also a population of study-specific research methods and conditions. This population of participants, methods, and conditions is randomly sampled each time the significance test is conducted (Fisher, 1922, p. 313).

Second, in scientific contexts, the population is not fully known. By definition, a “scientist” does not fully understand the relevant and irrelevant aspects of the populations that they are studying. Indeed, it is for this reason that they are studying those populations! Consequently, scientists face a reference class problem when attempting to specify the population to which their Type I error rate applies (Venn, 1876). They handle this problem by making theoretically informed guesses about the relevant and irrelevant aspects of the populations that are the subject of their statistical inferences. However, these guesses can be wrong! As Fisher (1956) explained, during significance testing “the population in question is hypothetical, … it could be defined in many ways, and … the first to come to mind may be quite misleading” (p. 78). Hence, scientists must continually ask themselves: “Of what population is this a random sample?” (Fisher, 1922, p. 313).

Third, Type I error rates are based on an imaginary situation in which random sampling error is the *only* source of error that can affect a researcher’s decision to reject the null hypothesis. Of course, in the real world, many other sources of error can influence a researcher’s decision (e.g., errors in data collection and entry, errors in research methodology, and/or errors in theoretical interpretation). However, Type I error rates do not refer to any of these other sources of error. They only refer to errors based on random sampling error (Berk et al., 1995, p. 423; Fisher, 1956, p. 44; Neyman & Pearson, 1928, pp. 177, 232). Hence, we must imagine that *if* a null hypothesis was true (i.e., if all samples were drawn from the null population), and *if* random sampling error was the *only* source of decision-making error, *then* the Type I error rate would indicate the frequency with which a researcher would reject the null hypothesis in a long run of repeated random sampling.

Finally, the idea of a frequency of decision-making errors during a long run of repeated random sampling from a null population is consistent with the Neyman-Pearson approach to hypothesis testing (Neyman & Pearson, 1928, 1933). However, a Type I error can also be conceptualised within the alternative Fisherian approach (Fisher, 1956, 1971). In this case, a Type I error *probability* (not rate) represents the *epistemic* probability (not aleatory probability) of making a Type I error in relation to a *single* decision (not a long run of decisions) to reject a null hypothesis based on an observed test statistic from the *current sample* (not a long run of samples; Rubin, 2020b, 2021b).

## What Is Type I Error Rate Inflation?

Type I error rate inflation occurs when the *actual* Type I error rate is higher than the *nominal* Type I error rate. The nominal Type I error rate is the rate that is set by the researcher, and it is used to determine whether an observed $p$ value is "significant" or "nonsignificant." Hence, the nominal Type I error rate is also referred to as a *significance threshold* or *alpha level*. It is used to control the frequency of making a Type I error during a long run of repeated sampling.

The actual Type I error rate can be higher than the nominal Type I error rate in the context of multiple testing. For example, imagine that a researcher aims to make a decision about a null hypothesis based on three tests of that hypothesis, each with an alpha level of .050. Further imagine that the researcher is prepared to accept a significant result on *any* of the three tests as sufficient grounds for rejecting the null hypothesis. In this case, the researcher's actual (familywise) Type I error rate for their decision will be .143. Consequently, if they set the alpha level for their decision at .050, then their actual Type I error rate (.143) will be inflated above their nominal Type I error rate (.050). As I discuss later, this multiple testing problem underlies several of the research practices that are thought to inflate Type I error rates (selective inference, forking paths, exploratory analyses, *p*-hacking, optional stopping).

Importantly, the word "actual" in the phrase "actual Type I error rate" does not imply that we are able to identify "real" false positive results in any given study. The probability of a "real" false positive result would refer to the conditional posterior probability that a null hypothesis is true given its rejection (i.e., Pr[$H_0$ is true | reject $H_0$]). However, it is not possible to quantify this probability in scientific contexts (Meehl, 1997, p. 397; Neyman & Pearson, 1928, p. 176; Pollard & Richardson, 1987, p. 162). Instead, we must consider the hypothetical probability of rejecting a null hypothesis when it is true (i.e., Pr[reject $H_0$ ; $H_0$ is true]).[1] It is this hypothetical probability, rather than the conditional posterior probability, that represents the actual Type I error rate.

People sometimes confuse the actual Type I error rate with the conditional posterior probability (Mayo & Morey, 2017; Pollard & Richardson, 1987). For example, they might argue that, if you reject 200 null hypotheses using an alpha level of .050, and only 100 of those hypotheses are true, then you will end up with five actual Type I errors (i.e., 100 × .050). However, this scenario refers to the probability that a null hypothesis is true when it is rejected, Pr($H_0$ is true | reject $H_0$), rather than the probability of rejecting a null hypothesis when it is true, Pr(reject $H_0$; $H_0$ is true). Confusing these two types of probability can be described as a Bayesian inversion fallacy (Gigerenzer, 2018; Greenland et al., 2016). During significance testing, the actual Type I error rate does not refer to the probability or prevalence of true null hypotheses; it simply assumes that each null hypothesis is true and then represents the frequency with which each hypothesis would be rejected given random sampling error per se (Fisher, 1971, p. 17). Hence, to return to the previous example, a person who rejects 200 null hypotheses using an alpha level of .050 should expect to make 10 *actual* Type I errors (i.e., 200 × .050), not 5 (100 × .050), because they should assume (imagine) that all 200 hypotheses are true.[2]

Finally, it should be noted that nominal Type I error rates tend to be based on a research field's conventional alpha level (e.g., $p \leq .050$). Nonetheless, individual researchers can choose alpha levels that are more or less stringent than the conventional level (Benjamin et al., 2018; Maier & Lakens, 2022). Hence, Type I error rate inflation should be distinguished from unconventional alpha levels. In particular, Type I error rate inflation should be judged by comparing the actual Type I error rate for a statistical inference with the nominal Type I error rate for that inference (i.e., the alpha level) on the understanding that the nominal level may be set at a conventional or unconventional level. For example, although an actual Type I error rate of .100 may be higher than a field's conventional alpha level of .050, it cannot be said to be "inflated" if the researcher has explicitly set their alpha level at the unconventional level of .100. If Type I error rate inflation was judged relative to the conventional alpha level, rather than the nominal alpha level, then any researcher who set their alpha level higher than the conventional level could be said to have an inflated Type I error rate!

## Type I Error Rates Do Not Refer to Theoretical Errors

It is important to distinguish between *statistical inferences* and *theoretical inferences* because Type I error rates only refer to the former (e.g., Meehl, 1978, p. 824; Meehl, 1997, p. 401; see also Bolles, 1962; Chow, 1998; Cox, 1958, p. 357; Hager, 2013, p. 259; Mayo & Spanos, 2006, p. 341; Neyman, 1950, p. 290). During significance testing, a statistical inference refers to a statistical null hypothesis which states that samples are drawn from a study-specific null population in the context of random sampling error per se. Statistical inferences are always supported by inferential statistics, and they usually describe test results as being either "significant" or "nonsignificant." For example, the following statement is a statistical inference: "Compared to male participants, female participants reported significantly more positive attitudes towards ice cream, $t(326) = 2.62$, $p = .009$." In this example, the researcher has provisionally rejected the statistical null hypothesis that female participants *do not* report more positive attitudes towards ice cream than male participants. Their alpha level (e.g., .050) indicates the frequency with which they would make an error in rejecting this statistical null hypothesis in a long run of random sampling from the statistical null population.

In contrast to statistical inferences, theoretical inferences refer to substantive hypotheses that generalize beyond the specifics of the current study. Consequently, they are not directly associated with study-specific Type I error rates. For example, a substantive inference might be that, "compared to men, women have more positive attitudes towards ice cream."

The distinction between statistical and theoretical inferences leads to a parallel distinction between statistical and theoretical errors. Statistical errors refer to Type I and Type II errors. In contrast, theoretical errors refer to a wide range of misinterpretations about (a) theory (e.g., misinterpreted theoretical rationales, hypotheses, and predictions), (b) methodology (e.g., misspecified participant populations, sampling procedures, testing conditions, stimuli, manipulations, measures, controls, etc.), (c) data (e.g., misspecified procedures for data selection, entry, coding, cleaning, aggregation, etc.), and (d) analyses (e.g., misspecified statistical models and assumptions, misinterpreted statistical results).

Theoretical errors may occur in the absence of statistical errors. In other words, researchers may make theoretical misinterpretations before and after correctly rejecting statistical null hypotheses. Meehl's (1990b, 1997) concept of *crud* provides a good example. Crud is a real but theoretically trivial effect (e.g., a methodological artefact). As Meehl (1990b) explained, crud consists of "*real* differences, *real* correlations, *real* trends and patterns" (pp. 207-208, emphasis in original). Hence, crud "does not refer to statistical error, whether of the first or the second kind" (i.e., Type I or II errors; Meehl, 1997, p. 402). In particular, "we are not dealing here with some source of statistical error (the occurrence of random sampling fluctuations). That source of error is limited by the significance level we choose" (Meehl, 1990b, p. 207). Nonetheless, researchers may make theoretical errors about crud by misinterpreting it as theoretically important effects.

Such errors may be conceptualized as *theoretical* false positives (i.e., incorrectly accepting crud as being supportive of a substantive alternative hypothesis) rather than *statistical* false positives (i.e., incorrectly rejecting a statistical null hypothesis).

Theoretical errors may also have a larger impact than statistical errors (Bolles, 1962, p. 645; Cox, 1958, p. 357; Fisher, 1926, pp. 504-505; Greenland, 2017b, p. 640). Hence, a researcher's probability of incorrectly rejecting a substantive null hypothesis and incorrectly accepting a substantive alternative hypothesis may be greater than their alpha level because their decisions are influenced by numerous theoretical errors in addition to Type I errors.

Researchers may sometimes confuse statistical errors with theoretical errors and assume that their Type I error rate indicates the probability of incorrectly rejecting a substantive null hypothesis in the real world rather than a statistical null hypothesis in an imaginary long run of repeated sampling. Greenland (2017b, 2023) described this confusion as "statistical reification." He argued that researchers sometimes forget that their "statistical analyses are merely thought experiments" based on idealized assumptions that are unlikely to be true in the real world. The outcome of this confusion is "overconfident inference" (Greenland, 2017a, p. 4; Greenland, 2017b, p. 640; see also Brower, 1949, p. 327; Gigerenzer, 1993, p. 329; Meehl, 1990b, p. 225). In particular, researchers may have unwarranted credulity in a substantive conclusion based on their incorrect belief that the Type I error rate covers one or more theoretical errors.

Finally, unlike statistical errors, theoretical errors cannot be quantified. As Meehl (1997) explained, "it is tempting to conflate the inference relation between statistics and parameters with the relation between accepted parameter values and the substantive theory; and because the former is numerified (e.g., a Bayesian posterior, a confidence belt, a significance level), one tends to think the latter is numerified also, or (somehow) *should* be" (p. 397; emphasis in original). However, as Neyman and Pearson (1928) explained, "the sum total of the reasons which will weigh with the investigator in accepting or rejecting the hypothesis can very rarely be expressed in numerical terms. All that is possible for him is to balance the results of a mathematical summary, formed upon certain assumptions, against other less precise impressions based upon a priori or a posteriori considerations" (p. 176). Note that rigorous pretesting, validation, and preregistration of a priori considerations and testing assumptions do not make them infallible and do not exclude their contribution to the overall probability of making a substantive error in scientific research (Uygun-Tunç & Tunç, 2023, p. 6). Even in the most rigorous research, the probability of a substantive error refers to not only statistical errors but also the unquantifiable and unknowable probability that the researcher has made a theoretical error.

# The Impact of Various Research Practices on Type I Error Rates

## Model Misspecification

Model misspecification does not inflate Type I error rates because a Type I error rate assumes that the associated null model is "true" or at least "adequate," which means that it is correctly (adequately) specified, and that the only source of influential error is random sampling error. To argue that modelling error inflates Type I error rates is to commit the Bayesian inversion fallacy and believe that Type I error rates are influenced by the probability of the correctness of the model to which they refer (Gigerenzer, 2018; Greenland et al., 2016; Pollard & Richardson, 1987).

This is not to say that null models are always correctly specified. The point here is only that the frequentist concept of a Type I error rate assumes that they are. Of course, in reality, null models may be misspecified. In particular, (a) the statistical null model may not adequately represent the experimental null model, and/or (b) the experimental null model may not adequately represent the theoretical null model (Devezer & Buzbas, 2023; Spanos, 2006). These model misspecifications may then lead to serious inferential errors. However, these errors are theoretical rather than statistical. Hence, it is more appropriate to conceive model misspecification as inflating *Type III* errors, rather than Type I errors. As Dennis et al. (2019) explained, Type III errors occur when "neither the null nor the alternative hypothesis model adequately describes the data (Mosteller, 1948)" (p. 2).

## Multiple Testing

The term *multiple testing* covers several types of testing situation. Here, I distinguish between (a) single tests of multiple individual hypotheses and (b) multiple tests of a single joint hypothesis. I argue that Type I error rate inflation never occurs in the first situation, and that it is not problematic in the second situation because it is easily identified and resolved.

### *Single Tests of Multiple Individual Null Hypotheses*

Imagine that a researcher conducts a study in which they test for gender differences on 20 different dependent variables that measure a variety of different attitudes (e.g., attitudes towards abortion, environmentalism, the death penalty, pizza, ice cream, and so on). In this case, the researcher is testing 20 different null hypotheses (i.e., $H_{0,1}, H_{0,2}, H_{0,3}, ,\ldots H_{0,20}$). Further imagine that the researcher sets their alpha level at the conventional level of .050 for each statistical inference that they make about each null hypothesis. We can call this alpha level the *individual* alpha level (i.e., $\alpha_{\text{Individual}} = .050$) because it refers to the frequency with which the researcher would incorrectly reject each individual null hypothesis. This type of testing situation represents single tests of multiple individual null hypotheses. In this case, there is no more than one opportunity to make a Type I error in relation to each individual null hypothesis because none of the hypotheses undergo more than one test.

In a review of 109 research articles published in the journals *Behavior Research Methods* and *Psychological Science* between 2021 and June 2022, García-Pérez (2023) found that *all* cases of multiple testing represented single tests of multiple individual hypotheses. Extrapolating from this evidence, it is likely that single tests of multiple individual hypotheses represent the most common type of multiple testing.

It has also been repeatedly shown that single tests of multiple individual hypotheses do not result in Type I error rate inflation, regardless of how many single tests are undertaken (Armstrong, 2014, p. 505; Cook & Farewell, 1996, pp. 96–97; Fisher, 1971, p. 206; García-Pérez, 2023, p. 15; Greenland, 2021, p. 5; Hewes, 2003, p. 450; Hitchcock & Sober, 2004, pp. 24-25; Hurlbert & Lombardi, 2012, p. 30; Matsunaga, 2007, p. 255; Molloy et al., 2022, p. 2; Parker & Weir, 2020, p. 564; Parker & Weir, 2022, p. 2; Rothman, 1990, p. 45; Rubin, 2017b, pp. 271–272; Rubin, 2020a, p. 380; Rubin, 2021a, 2021c, pp. 10978-10983; Rubin, 2024, p. 3; Savitz & Olshan, 1995, p. 906; Senn, 2007, pp. 150-151; Sinclair et al., 2013, p. 19; Tukey, 1953, p. 82; Turkheimer et al., 2004, p. 727; Veazie, 2006, p. 809; Wilson, 1962, p. 299). Consequently, it can be concluded the most common form of multiple testing does not inflate Type I error rates.

People sometimes doubt the absence of Type I error rate inflation during single tests of multiple individual null hypotheses, but it is easy to demonstrate: In general, the actual Type I error rate is computed using the formula $1 - (1 - \alpha)^k$, where $k$ is the number of tests that are used to make a decision about a specific null hypothesis. During single tests of multiple individual null hypotheses, $k = 1$ because only one test is used to make a decision about each null hypothesis.

Hence, the actual Type I error rate for each statistical inference is equal to $1 - (1 - \alpha_{\text{Individual}})^1$, which is equal to the nominal $\alpha_{\text{Individual}}$ (e.g., $1 – [1 - .050]^1 = .050$; see Rubin, 2024, Confusion II).

### ***Multiple Tests of a Single Joint Null Hypothesis***

Now imagine that the researcher groups some of the 20 dependent variables together for some reason. For example, they might consider attitudes about abortion, environmentalism, and the death penalty to be theoretically exchangeable in the context of a broader joint hypothesis about political orientation. In this case, the researcher might be prepared to accept a significant gender difference in relation to *at least one* of these three hypotheses in order to make a statistical inference that there is a significant gender difference in political orientation. Here, the three null hypotheses ($H_{0,1}$, $H_{0,2}$, & $H_{0,3}$) are treated as *constituent* null hypotheses that form part of a broader *joint* intersection null hypothesis about political orientation: "$H_{0,1}$ and $H_{0,2}$ and $H_{0,3}$." In this case, the rejection of *any one* of the three constituent null hypotheses is sufficient to reject the entire intersection null hypothesis and make the statistical inference that, for example, "compared to male participants, female participants reported significantly more left-wing attitudes: abortion $t(326) = 2.54$, $p = .011$; environmentalism $t(326) = .030$, $p = .979$; death penalty $t(326) = 1.44$, $p = .150$." In this example, the test statistics refer to each of the tests of the three constituent hypotheses and, because at least one of the $p$ values is significant at the conventional level ($p = .011$), the researcher can reject the entire joint null hypothesis about "left-wing attitudes."

During this *union-intersection testing* (e.g., Hochberg & Tamrane, 1987, p. 28; Kim et al., 2004), the actual familywise Type I error rate for the joint null hypothesis is always larger than the nominal alpha level for each of the constituent hypotheses: $\alpha_{\text{Constituent}}$. For example, if $\alpha_{\text{Constituent}}$ is set at .050, then the familywise error rate will be $1 - (1 - \alpha_{\text{Constituent}})^k$, where $k$ is the number of constituent null hypotheses that are included in the joint null hypothesis. Hence, in the present example, the familywise error rate will be $1.00 - (1 - .050)^3 = .143$.

One concern here is that the Type I error rate for decisions about each of the *constituent* null hypotheses becomes inflated. However, this concern is unwarranted. $\alpha_{\text{Constituent}}$ is the nominal alpha level for the per comparison Type I error rate, and this error rate does not become inflated during multiple tests of a single joint null hypothesis for the same reason that $\alpha_{\text{Individual}}$ does not become inflated during single tests of multiple individual null hypotheses (i.e., $1 - [1 - \alpha_{\text{Constituent}}]^1 = \alpha_{\text{Constituent}}$; Rubin, 2021c, p. 10979; Rubin, 2024; Tukey, 1953).

Another concern is that the Type I error rate for the decision about the *joint* null hypothesis can become inflated. This concern is legitimate. In order to identify this Type I error rate inflation, we need to check whether the *actual* familywise Type I error rate for the *joint* null hypothesis is higher than the *nominal* familywise Type I error rate for the *joint* null hypothesis: $\alpha_{\text{Joint}}$. If both $\alpha_{\text{Constituent}}$ and $\alpha_{\text{Joint}}$ are set at the same level (e.g., the conventional level of .050), then the actual Type I error rate for the joint null hypothesis will be inflated above $\alpha_{\text{Joint}}$. However, researchers can avoid this inflation by adjusting $\alpha_{\text{Constituent}}$ downwards until the familywise Type I error rate is at $\alpha_{\text{Joint}}$. For example, if $k = 3$ and both $\alpha_{\text{Constituent}}$ and $\alpha_{\text{Joint}}$ are originally set at .050, then a Bonferroni adjustment (i.e., $\alpha \div k$) can be used to reduce $\alpha_{\text{Constituent}}$ from .050 to .017 in order to maintain the familywise Type I error rate for the joint null hypothesis at the nominal $\alpha_{\text{Joint}}$ of .050.

What happens if researchers do not adjust $\alpha_{\text{Constituent}}$? In this case, there will be Type I error rate inflation. However, this inflation can be easily identified and resolved by readers. For example, reconsider the previous statistical inference: "Compared to male participants, female participants reported significantly more left-wing attitudes: abortion $t(326) = 2.54$, $p = .011$; environmentalism $t(326) = .030$, $p = .979$; death penalty $t(326) = 1.44$, $p = .150$." Here, it is clear that three test results are used to make a single statistical inference about a joint hypothesis that is broader than any of the three constituent hypotheses (i.e., “significantly more left-wing attitudes”). In the absence of any other information, we can assume that the tests use a conventional unadjusted $\alpha_{Constituent}$ of .050. It is also clear that not all of the tests need to be significant to make the statistical inference (i.e., *p*s = .011, .979, & .150). Hence, the statistical inference is based on union-intersection testing. Finally, we can assume that $\alpha_{Joint}$ has also been set at the conventional level of .050. Consequently, we can conclude that the actual familywise Type I error rate for this statistical inference is greater than a conventional $\alpha_{Joint}$ of .050. In other words, there is Type I error rate inflation. However, the inflation is transparent, and so it can be the target of criticism by reviewers and readers. The inflation is also easily computed (.143 relative to a conventional $\alpha_{Joint}$ of .050). Finally, the inflation is easily resolved. For example, any reader can implement a Bonferroni adjustment to remove the Type I error rate inflation and conclude that, using a conventional $\alpha_{Joint}$ of .050 and an adjusted $\alpha_{Constituent}$ of .017, the researcher’s overall statistical inference would remain valid due to the $p = .011$ result.

Is transparent reporting necessary to identify and resolve Type I error rate inflation in this situation? In particular, to determine *k*, do readers need to be aware of all other tests that a researcher conducted? No, they do not, because the researcher is obviously not using any other test results to make their statistical inference. More generally, *k* is the number of tests that are formally associated with a reported statistical inference. *k* is not necessarily the number of tests that a researcher conducted, or planned to conduct, in their study. This position is consistent with philosophies of science and statistics that focus on the logical relations between a hypothesis and its evidence without considering their psychological origins (Fisher, 1956; Popper, 1962, 2002). Following this philosophy, a familywise error rate can be logically deduced from a potentially unplanned reported statistical inference rather than needing to be predetermined based on planned inferences that may never eventuate (cf. Mayo, 1996, p. 296). Furthermore, undisclosed alterations to planned inferences do not impact the validity of a familywise error rate that is logically deduced from an unplanned statistical inference.

Certainly, a familywise error rate may be logically inconsistent with a reported statistical inference (Rubin, 2024). For example, a reported statistical inference may be formally associated with *three* significance tests when the familywise error rate is computed on the basis of only *two* of these tests. However, as previously discussed, this logical error is readily identifiable by checking (a) the number of tests that are reported as being associated with a statistical inference and (b) the number of tests that are reported as being used to compute the associated familywise error rate. Again, it is not necessary to consult a preregistered plan to identify and resolve this logical inconsistency in the reported information.

Finally, it is worth noting that flexible theorizing and flexible theories allow the post hoc construction of multiple different joint null hypotheses and associated familywise error rates. As I discuss later in the section on *p*-hacking, researchers’ personal biases (including hindsight and confirmation biases) may influence which constructions are tested and reported. However, scientific hypotheses are usually judged on the basis of theoretical virtues (e.g., plausibility, universality, precision, depth, breadth, coherence, parsimony, etc.; Kuhn, 1977, p. 103; Mackonis, 2013; Popper, 1962, p. 56, p. 232; Popper, 2002, p. 438) rather than a researcher’s personal motives and biases (e.g., to try to report a significant result and publish a paper; Reichenbach, 1938, p. 5; Popper, 1962, p. 140; Popper, 2002 p. 7; Rubin, 2022, pp. 541-542; Rubin & Donkin, 2022, p. 19). Consequently, the lack of transparency about these personal biases does not prevent a rigorous appraisal of the theoretical rationales in question (Rubin, 2017, p. 314; Rubin, 2022, p. 539).

***Studywise Type I Error Rates***

Researchers are sometimes concerned about the probability of making at least one Type I error in their study or experiment. This concern about a *studywise* or *experimentwise* Type I error rate implies that they are making a statistical inference about a joint studywise null hypothesis that can be rejected by *any* single significant result in their study. In practice, however, it is not common for researchers to make this type of inference because joint studywise null hypotheses do not usually have a useful theoretical basis. Consequently, most researchers should not be concerned about their studywise Type I error rate because it relates to a joint null hypothesis that they are not testing. Instead, researchers should be concerned about the error rates for the individual and/or joint hypotheses about which they actually make statistical inferences (Rubin, 2021c, p. 10991; Rubin, 2024).

To illustrate, reconsider the previous example in which a researcher tested for gender differences on 20 different dependent variables that measured a wide range of attitudes. Recall that three of these attitudes could be grouped into a theoretically meaningful joint null hypothesis about political orientation (i.e., attitudes towards abortion, environmentalism, and the death penalty). However, the other attitudes had nothing to do with political orientation and so had no theoretical basis for being included in this joint null hypothesis (e.g., attitudes about pizza and ice cream). The same issue of theoretical relevance applies to the consideration of joint studywise null hypotheses. Hence, in the present example, the researcher may not have a good theoretical basis for considering all 20 null hypotheses as constituents of a single joint null hypothesis that can be rejected following at least one significant gender difference. This being the case, they should not make a statistical inference about this studywise null hypothesis, and they should not be concerned about its associated studywise Type I error rate.

To be clear, I am not claiming that studywise error rates are *always* irrelevant. They are relevant whenever researchers make statistical inferences about associated joint studywise null hypotheses on the basis of union-intersection testing and, in this case, researchers should adjust their $\alpha_{\text{Constituent}}$ in order to control their studywise error rate at $\alpha_{\text{Joint}}$. My point is that this situation is likely to be relatively rare because, typically, a single study will include a collection of disparate hypotheses that do not form a theoretically meaningful joint studywise null hypothesis in aggregate (for similar views, see Bender & Lange, 2001, p. 343; Hancock & Klockars, 1996, p. 270; Hewes, 2003, p. 450; Hochberg & Tamrane, 1987, p. 7; Morgan, 2007, p. 34; Oberauer & Lewandowsky, 2019, p. 1609; Parker & Weir, 2020, p. 2; Perneger, 1998, p. 1236; Rothman et al., 2008, pp. 236-237; Rubin, 2017b, p. 271; Rubin, 2020a, p. 382; Schulz & Grimes, 2005, p. 1592). In such cases, a statistical inference about a joint studywise null hypothesis would need to be relatively vague and atheoretical, along the lines of: "The study's effect was significant." It is not common for researchers to make this sort of abstract atheoretical statistical inference. Instead, researchers usually make more specific, theory-based statistical inferences that relate to substantive theoretical claims. Hence, as in the gender differences example, the researcher might infer that, "compared to male participants, female participants reported significantly more left-wing attitudes: abortion $t(326) = 2.54$, $p = .011$; environmentalism $t(326) = .030$, $p = .979$; death penalty $t(326) = 1.44$, $p = .150$."

***Summary***

Type I error rate inflation is neither common nor problematic during multiple testing. The most common form of multiple testing is represented by single tests of multiple individual null hypotheses (García-Pérez, 2023). Most experts agree that this type of multiple testing does not inflate relevant Type I error rates because only one test is used to make a decision about each null hypothesis. Consequently, the most common form of multiple testing does not inflate Type I error rates. Indeed, the inappropriate (illogical) use of alpha adjustments in this type of testing situation may result in Type I error rate *deflation,* rather than *inflation* (Rubin, 2024).

A less common form of multiple testing is represented by multiple tests of a single joint null hypothesis (union-intersection testing). Type I error rate inflation is possible in this case because, if both $\alpha_{\text{Constituent}}$ and $\alpha_{\text{Joint}}$ are set at the same level, then the actual Type I error rate for the joint null hypothesis will be inflated above $\alpha_{\text{Joint}}$. However, this inflation can be easily identified by comparing (a) the number of statistical tests that are formally associated with a reported statistical inference and (b) the number of statistical tests that are used to compute the associated familywise error rate. Type I error rate inflation will be readily apparent if (a) is larger than (b), and it can be easily resolved by recomputing the familywise error rate using the correct number of tests as represented by (a).

Finally, multiple testing increases the studywise Type I error rate above $\alpha_{\text{Constituent}}$. However, researchers do not usually make statistical inferences about the associated joint studywise null hypotheses, and so this increase is usually irrelevant. Nonetheless, if it does become relevant, then it can be easily identified and resolved by adjusting $\alpha_{\text{Constituent}}$ as described above.

## Selective Inference

Imagine that a researcher checks the effect sizes of 100 correlations between 200 different variables (e.g., $x_1$–$y_1$, $x_2$–$y_2$, $x_3$–$y_3$, … $x_{100}$–$y_{100}$) and then decides to perform a significance test on the correlation between variables $x_{57}$ and $y_{57}$ because it had the largest effect size (for a similar example, see Taylor & Tibshirani, 2015, p. 7629). If the researcher uses the significant $x_{57}$–$y_{57}$ result to reject a joint null hypothesis that could be rejected by a significant result on *any* of the 100 constituent correlations, then they would need to adjust their $\alpha_{\text{Constituent}}$ downwards to prevent their familywise error rate from exceeding their $\alpha_{\text{Joint}}$. However, it is uncommon for researchers to make such a *selective inference*. Instead, researchers tend to make statistical inferences that are *limited to* their selected test and data rather than extended to unselected tests and data that they could have used (Birnbaum, 1962, pp. 278-279; Cox, 1958, p. 359-361; Cox & Mayo, 2010, p. 296; Lehmann, 1993, pp. 1245-1246; Mayo, 2014, p. 232; Reid & Cox, 2015, p. 300). Hence, the researcher in the present example would be more likely to use the significant $x_{57}$–$y_{57}$ result to reject an individual null hypothesis about the relationship between $x_{57}$ and $y_{57}$ rather than a joint intersection null hypothesis about the relationships between $x_1$–$y_1$, $x_2$–$y_2$, $x_3$–$y_3$, … $x_{100}$–$y_{100}$. In the individual case, $\alpha_{\text{Constituent}}$ and $\alpha_{\text{Joint}}$ are irrelevant to the researcher's inference, and an unadjusted $\alpha_{\text{Individual}}$ can be used without any concern about Type I error rate inflation. Note that, as discussed previously, this point stands even if the researcher does not disclose that they considered the other 99 correlations because those correlations are not formally associated with their reported statistical inference, which is about the relationship between $x_{57}$ and $y_{57}$.

Alternatively, the researcher may decide to make a statistical inference about a joint null hypothesis based on a family of correlations, such as "$x_{56}$–$y_{56}$ and $x_{57}$–$y_{57}$ and $x_{58}$–$y_{58}$." For example, in an fMRI study, these correlations might occur within the same brain region and, again, they may be selected because they "look promising." In this case, the familywise error rate can be logically deduced from the statistical inference (i.e., $k = 3$), and the formal theoretical rationale for the joint hypothesis can be critically evaluated on the basis of its theoretical virtues. Again, the familywise error rate for the statistical inference is not inflated by the existence of alternative planned or unplanned statistical inferences that could have been based on other joint null hypotheses from the set of correlations.

People sometimes confuse individual and constituent null hypotheses in this situation. For example, it is true that $\alpha_{Constituent}$ would need to be adjusted when using the significant $x_{57}$–$y_{57}$ correlation to reject a joint null hypothesis that could be rejected by at least one significant result from among 100 potential correlations, because $k = 100$ for this selective inference. Similarly, as described above, $\alpha_{Constituent}$ would need to be adjusted when using the significant $x_{57}$–$y_{57}$ correlation to reject a joint null hypothesis that could be rejected by at least one significant result from among three potential correlations (e.g., "$x_{56}$–$y_{56}$ and $x_{57}$–$y_{57}$ and $x_{58}$–$y_{58}$"), because $k = 3$ for this selective inference. However, this point does not imply that $\alpha_{Individual}$ needs to be adjusted when using the significant $x_{57}$–$y_{57}$ result to reject the $x_{57}$–$y_{57}$ null hypothesis, because $k = 1$ for this individual inference.

Part of the confusion here is that people assume that selective *analyses* always lead to selective *inferences*. On the contrary, selective analyses usually lead to inferences that are *limited to the selection*. For example, a researcher may select data and/or variables for analysis from a broader set. However, this act does not then obligate them to make a selective inference about a joint null hypothesis that refers to other data or variables from that set. Instead, it would be more usual for them to limit their inference to the data or variables that they have selected (i.e., an *unconditional* inference about the selected data rather than a *conditional* inference that is conditioned on the selection procedure).

More generally, it is important to avoid confusion about the reference sets to which statistical inferences refer. As Neyman (1950) explained, "many errors in computing probabilities are committed because of losing sight of the set of objects to which a given probability is meant to refer" (p. 15). A Type I error rate is meant to refer to a decision about a specified statistical (individual or joint) null population and not to any broader population from which that null population may have been selected. Hence, the selection of a particular subset of data for testing from a more inclusive set because it "looks promising" will not inflate the Type I error rate for a statistical inference as long as that inference refers to a population based on *that particular subset of data* and not to a population based on the more inclusive set of data (for related discussions, see Fisher, 1956, p. 88-89; Kotzen, 2013, p. 167; Rubin, 2021c, p. 10983). For example, conducting multiple tests of a gender difference in the preference for ice cream in different European countries and then selectively reporting the results for France may bias claims about a gender difference in Europe but not claims about a gender difference in France. Of course, the result in France may represent a Type I error. However, this possibility is covered by the alpha level of the test that was conducted on the French data.

## Forking Paths

A forking path occurs during data analysis when a result from one sample of data inspires a researcher to conduct a specific test in a situation in which they would have conducted a different test if they had observed a different result using a different sample (Gelman & Loken, 2014). For example, a researcher might report that "this variable was included as a covariate in the analysis because it was significantly correlated with the outcome variable." The implication here is that the variable would *not* have been included as a covariate if it had *not* been significantly correlated with the outcome variable in a different sample of data. Consequently, if the researcher makes a statistical inference about a joint null hypothesis that can be rejected following a significant result on at least one of the two tests (i.e., the test that includes the covariate and the test that does not), then their familywise Type I error rate will be greater than the $\alpha_{Constituent}$ for each test (Rubin, 2020a, p. 380). Hence, the forking paths problem resolves to a case of multiple testing in which the "invisible multiplicity" is only apparent in a long run of repeated sampling (Gelman & Loken, 2014, p. 460).

The forking paths problem assumes that a researcher will make a statistical inference about a joint null hypothesis that comprises the two forking paths in their analysis (e.g., a test that includes a covariate and a test that does not). *If* the researcher makes this statistical inference, then they can adjust their $\alpha_{Constituent}$ to retain their $\alpha_{Joint}$ at the actual familywise error rate (i.e., $\alpha_{Constituent}$ ÷ 2; Rubin, 2017a, p. 324). However, it is more likely that the researcher would make a more limited statistical inference based on only *one* test in *one* of the two forking paths (Birnbaum, 1962, pp. 278-279; Cox, 1958, p. 359-361; Cox & Mayo, 2010, p. 296; Lehmann, 1993, pp. 1245-1246; Mayo, 2014, p. 232; Reid & Cox, 2015, p. 300). In this case, the researcher's inference would refer to an imaginary long run of repeated sampling that, for example, always includes the variable as a covariate and never excludes it. An unadjusted $\alpha_{Individual}$ would then be appropriate. Note that the Type I error rate is not inflated in either of these two situations.

## Exploratory Analyses

An exploratory data analysis is one in which a study's analytical approach is guided by idiosyncratic results in one sample of data that may not occur in other samples. Consequently, the tests that are undertaken in one instance of an exploratory analysis may be quite different to those undertaken in repetitions of that analysis. This issue leads to multiple tests of a joint studywise null hypothesis both *within each repetition* of the exploratory study and *across the long run of its repetitions*. In theory, the associated studywise error rate would then need to account for every null hypothesis that could possibly be tested during the exploratory analysis and its repetitions. This situation has led several people to conclude that the exploratory studywise error rate cannot be computed or controlled (e.g., Hochberg & Tamrane, 1987, p. 6; Nosek & Lakens, 2014, p. 138; Wagenmakers, 2016). As Wagenmakers (2016) explained, "the problem is one of multiple comparisons with the number of comparisons unknown (De Groot, 1956/2014)."

There are three problems with this line of reasoning. First, in practice, a researcher is *least* likely to make a statistical inference about a joint studywise null hypothesis during an *exploratory* analysis because, in this situation, they are least likely to have a satisfactory theoretical rationale for aggregating a potentially infinite number of result-contingent null hypotheses into a single joint studywise null hypothesis.

Second, if a researcher does make a statistical inference about an exploratory joint studywise null hypothesis, then it would need to be relatively abstract and atheoretical (e.g., "the study's effect was significant"). Again, it is not common for researchers to make this type of statistical inference. Instead, they are likely to make more specific theory-based statistical inferences that relate to substantive theoretical inferences.

Finally, if a researcher was to make a statistical inference about an exploratory joint studywise null hypothesis, then they would need to tie it to specific statistical results. They could not simply report that, "based on an unknown number of unspecified tests that could have been conducted, the study's effect was significant." They would need to specify the tests (actual and potential) and results (*p* values) that form the basis for their statistical inference. Of course, they would only be able to specify a finite number of tests and, consequently, it would be possible for them to specify *k* and adjust $\alpha_{Constituent}$ in order to prevent Type I error rate inflation. In other words, the act of specifying a statistical inference includes making known the statistical tests upon which it rests. It is worth noting that recent work on multiverse analyses and specification curve analyses demonstrates the feasibility of making known large numbers of diverse statistical tests during exploratory data analyses (Del Giudice & Gangestad, 2021; Simonsohn et al., 2020; Steegen et al., 2016).

In summary, in most cases, there is no need for researchers to be concerned about the inflation of an exploratory studywise Type I error rate because this error rate is irrelevant to the more limited and theoretically defined statistical inferences that they usually make. However, if researchers do proceed to make vague atheoretical statistical inferences about exploratory joint studywise hypotheses, then they will need to specify the tests involved, and so they will be able to specify $k$, adjust $\alpha_{\text{Constituent}}$, and control the associated studywise error rate at $\alpha_{\text{Joint}}$.

## *P*-Hacking

*P*-hacking is a questionable research practice that is intended to find and selectively report significant results. In their seminal article on "false positive psychology," Simmons et al. (2011) proposed that *p*-hacking inflates Type I error rates due to multiple testing. As they explained (p. 1359),

> it is common (and accepted practice) for researchers to explore various analytic alternatives, to search for a combination that yields "statistical significance," and to then report only what "worked." The problem, of course, is that the likelihood of at least one (of many) analyses producing a falsely positive finding at the 5% level is necessarily greater than 5%.

On this basis, Simmons et al. (2011, p. 1359) argued that "undisclosed flexibility in data collection and analysis allows presenting anything as significant."

A more formal analysis of Simmons et al.'s (2011) argument is as follows: If both $\alpha_{\text{Constituent}}$ and $\alpha_{\text{Joint}}$ are set at .050 during the union-intersection testing of an exploratory joint studywise null hypothesis, then the actual exploratory studywise error rate will exceed $\alpha_{\text{Joint}}$, and it will be easier to reject the exploratory joint studywise null hypothesis than it would be if the studywise error rate matched $\alpha_{\text{Joint}}$. Importantly, however, this situation does not allow researchers to present "anything as significant." It only allows researchers to present the *exploratory joint studywise alternative hypothesis* as significant and, given its atheoretical rationale, this hypothesis will be abstract and scientifically useless, akin to: "The study's effect was significant." Again, researchers tend to make more specific and theoretically informative statistical inferences based on (a) single tests of theory-based individual hypotheses and (b) multiple (union-intersection) tests of theory-based joint null hypotheses. Hence, I consider the implications of *p*-hacking in each of these two contexts below.

### *P-Hacking During Single Tests of Multiple Individual Null Hypotheses*

As explained previously, the actual Type I error rate for a single test of an individual null hypothesis remains at $\alpha_{\text{Individual}}$ even if (a) multiple such tests are conducted side-by-side within the same study, and (b) some of the tests are conducted but not reported (for related discussions, see Rubin, 2017a, 2020a, 2021c, 2024). Simmons et al. (2011) are correct that, when $k \geq 2$, familywise Type I error rates will be greater than $\alpha_{\text{Constituent}}$ and greater than .050 when $\alpha_{\text{Constituent}}$ is set at .050. However, neither of these points imply the inflation of Type I error rates for statistical inferences based on single tests of multiple individual null hypotheses. The argument that *p*-hacking causes Type I error rate inflation during single tests of multiple individual null hypotheses confuses statistical inferences about *individual* null hypotheses with statistical inferences about *joint* null hypotheses.

To illustrate, consider Simmons et al.'s (2011) demonstration of the impact of *p*-hacking on Type I error rates. Simmons et al. performed multiple tests on a real data set until they found a significant result that supported the outlandish theoretical inference that listening to the Beatles' song "When I'm Sixty-Four" makes people chronologically younger relative to listening to a control song (so-called "chronological rejuvenation"). The researchers then performed a series of simulations to compute the percentage of random samples in which there was at least one significant result in a family of, for example, "three *t* tests, one on each of two dependent variables and a third on the average of these two variables" (i.e., Situation A in Table 1 of Simmons et al., 2011, p. 1361). Using an $\alpha_{Constituent}$ of .050, an actual familywise Type I error rate of .095 was calculated. Note that, because the dependent variables were correlated with one another, this error rate is lower than the expected rate of .150 (Simmons et al., 2011, p. 1365, Note 3). Nonetheless, it is higher than the $\alpha_{Constituent}$ of .050. Consequently, the researchers concluded that "flexibility in analyzing two dependent variables (correlated at $r$ = .50) nearly doubles the probability of obtaining a false-positive finding" (p. 1361). This conclusion is correct. However, the "false-positive finding" in question relates to the incorrect rejection of a *joint* null hypothesis (i.e., that song condition has no effect on *any* of the three dependent variables), not an *individual* null hypothesis (e.g., that song condition has no effect on the first dependent variable), and the statistical inferences that are made in Simmons et al.'s demonstration are about *individual* null hypotheses, not *joint* null hypotheses. Hence, although Simmons et al.'s conclusion is correct, it is also irrelevant to the type of statistical inference that they consider.

To illustrate further, consider this part of a larger example that Simmons et al. (2011, p. 1364, bold text in original) used to demonstrate a fictitious researcher's selective reporting. In the following extract, the bolded text refers to a statistical inference that the researcher decided to report because it referred to a significant result, and the nonbolded text refers to a statistical inference that the researcher decided not to report because it referred to a nonsignificant result:

> **An ANCOVA revealed the predicted effect: According to their birth dates, people were nearly a year-and-a-half younger after listening to "When I'm Sixty-Four" (adjusted *M* = 20.1 years) rather than to "Kalimba" (adjusted *M* = 21.5 years), *F*(1, 17) = 4.92, *p* = .040.** Without controlling for father's age, the age difference was smaller and did not reach significance (*M*s = 20.3 and 21.2, respectively), *F*(1, 18) = 1.01, *p* = .33.

Importantly, in the above extract, both of the researcher's statistical inferences are based on single tests of individual null hypotheses, each underwritten by a separate statistical test and *p* value (i.e., $k = 1$ for each decision about each null hypothesis). Consequently, the actual Type I error rates for each statistical inference are consistent with a conventional $\alpha_{Individual}$ of .050. Furthermore, the fact that the second test is not reported has no impact on the actual Type I error rate of the first test, because the Type I error rate for the first statistical inference refers to an imaginary long run of random sampling in which father's age is *always* included as a covariate in the test.

Simmons et al.'s (2011) concern about Type I error rate inflation would only be warranted if their fictitious researcher made a statistical inference about a *joint* null hypothesis that referred to *both* the bolded and the nonbolded tests. Obviously, in the current example, the researcher does not report the nonbolded test, and so they do not make a statistical inference about an associated joint null hypothesis. However, even if the researcher reported both tests as indicated above, they would still not be making a statistical inference about a joint null hypothesis. They would be following the more common approach of making two statistical inferences about two individual null hypotheses. They would only make a statistical inference about a joint null hypothesis if they used two or more significance tests to make a single decision about a single (joint) null hypothesis (Rubin, 2024). To illustrate, the following example represents a case of Type I error rate inflation during a statistical inference about a joint null hypothesis (assuming that $\alpha_{Constituent}$ and $\alpha_{Joint}$ are both set at .050):

> An ANCOVA found that type of song had a significant effect on birth date (father's age included as a covariate, $F(1, 17) = 4.92$, $p = .040$; father's age excluded as a covariate, $F(1, 18) = 1.01$, $p = .330$.

Contrary to the present view, Nosek et al. (2018) argued that nontransparent selective reporting inflates the Type I error rate. Consequently, as they explained, "transparent reporting that 1 in 20 experiments or 1 in 20 analyses yielded a positive result will help researchers identify the one as a likely false positive" (p. 2603). However, like Simmons et al. (2011), this explanation confuses statistical inferences about individual hypotheses with statistical inferences about joint hypotheses. If researchers make a statistical inference based on a single test of an individual hypothesis, then their actual Type I error rate for this inference will be $\alpha_{Individual}$ regardless of whether they make 1, 20, or a million other statistical inferences and even if their statistical result for that inference is the only significant result that they obtain or report. On the other hand, if researchers make a statistical inference based on the union-intersection testing of a joint null hypothesis that includes 20 constituent experiments or analyses, then their actual familywise Type I error rate for that inference will be $1 - (1 - \alpha_{Constituent})^{20}$. However, contrary to Nosek et al., this familywise error rate does not refer to a "likely false positive" in relation to any "one" experiment or analysis (i.e., a constituent hypothesis; Rubin, 2024). It refers to an incorrect decision about a *joint* null hypothesis that is based on the entire *family* of 20 experiments or analyses. Again, (a) if researchers do not make a statistical inference about this joint null hypothesis, then there is no need for them to be concerned about its familywise error rate, (b) researchers do not usually make statistical inferences about joint null hypotheses unless they have some theoretical rationale for doing so, and (c) if researchers do make statistical inferences about joint null hypotheses, then they can adjust their $\alpha_{Constituent}$ in order to control their $\alpha_{Joint}$ at some specified level.

***P-Hacking During Union-Intersection Testing of a Joint Null Hypothesis***

Does *p*-hacking inflate the actual familywise Type I error rate when a researcher makes a statistical inference about a joint null hypothesis based on union-intersection testing? No, it does not, because the familywise error rate refers to the researcher's specified reported statistical inference and not to any other statistical inference that they made and then failed to report during their *p*-hacking.

To illustrate, imagine that a researcher undertakes 20 union-intersection tests of a joint null hypothesis. They set $\alpha_{Constituent}$ at .0025 in order to maintain $\alpha_{Joint}$ at the conventional level of .050 (i.e., .0025 × 20). Using this $\alpha_{Constituent}$, they find no significant result. However, they notice that the smallest of their 20 *p* values is .004. They decide not to report their first set of union-intersection tests because it failed to reject the joint null hypothesis (i.e., they engage in selective reporting). Instead, they conduct a second set of union-intersection tests that includes 10 of the previous 20 null hypotheses, and they deliberately include the hypothesis that yielded the $p = .004$ result in this family of constituent hypotheses. This time, they set $\alpha_{Constituent}$ at .005, which continues to maintain $\alpha_{Joint}$ at the conventional level of .050 (i.e., .005 × 10). Note that the researcher now knows that the $p = .004$ result will be significant using an $\alpha_{Constituent}$ of .005, and so they know in advance that they are able to reject their new joint null hypothesis. Does this *p*- hacking and selective reporting inflate the actual familywise Type I error rate above $\alpha_{Joint}$? No, it does not, because the nominal $\alpha_{Joint}$ of .050 matches the actual familywise Type I error rate for the researcher's specified statistical inference, which has a *k* of 10 and an $\alpha_{Constituent}$ of .005.

The fact that *k* was 20 for an unreported statistical inference does not affect the Type I error rate for the reported statistical inference, for which *k* is 10. More generally, the Type I error rate for a statistical inference about an individual or joint null hypothesis is not impacted by other statistical inferences that could, would, or should have been made about other individual or joint null hypotheses. Additionally, it is not impacted by other statistical inferences that were planned or actually made and then either reported or not reported. Arguing that Type I error rates should be adjusted when making multiple statistical inferences confuses $\alpha_{Constituent}$ with $\alpha_{Individual}$ and $\alpha_{Joint}$. It is necessary to adjust $\alpha_{Constituent}$ when using multiple tests to make a single statistical inference about a single joint null hypothesis. However, it is not necessary to adjust either $\alpha_{Individual}$ or $\alpha_{Joint}$ when making multiple statistical inferences about multiple individual or joint null hypotheses (see Rubin, 2024, Confusion III).

Also, note that the researcher deliberately selected the reported set of union-intersection tests *because* they yielded a significant result. Again, however, this point does not alter the actual Type I error rate for their statistical inference, which matches their $\alpha_{Joint}$ of .050. Consequently, they are entitled to claim that their test provides a conventionally low rate of erroneous rejection in a hypothetical long run of repeated random sampling.

Finally, in reporting their test, the researcher is likely to provide a theoretical rationale for the inclusion and exclusion of the specific constituent hypotheses in their joint null hypothesis, and they may omit the fact the $p = .004$ result inspired their construction of this hypothesis. Again, however, this situation is not necessarily problematic because a researcher's personal motives and informal inspirations are not usually taken into account during the formal evaluation of scientific hypotheses (Popper, 1962, p. 140; Popper, 2002 p. 7; Reichenbach, 1938, p. 5; Rubin, 2022, pp. 541-542; Rubin & Donkin, 2022, p. 19). Instead, hypotheses tend to be judged on the basis of theoretical virtues (e.g., plausibility, universality, precision, depth, breadth, coherence, parsimony, etc.; Kuhn, 1977, p. 103; Mackonis, 2013; Popper, 1962, p. 56, p. 232; Popper, 2002, p. 438). Hence, in the present example, reviewers and other readers would be able to evaluate the quality of the researcher's theoretical rationale for their joint null hypothesis, even if they are unaware of the researcher's *p*-hacking (Rubin, 2017, p. 314; Rubin, 2022, p. 539). If the theoretical rationale for including and excluding the various constituent hypotheses in the joint hypothesis is cogently deduced from a well-established, coherent theory that explains a broad range of other effects in a relatively deep and efficient manner, then it should be given serious consideration regardless of the timing of its deduction.

***Summary***

In summary, it is true that *p*-hacking makes it easier to reject an exploratory joint studywise null hypothesis comprised of every null hypothesis that could possibly be tested in a study and its repetitions. However, researchers do not usually make statistical inferences about such hypotheses because they are not usually theoretically informative. For example, researchers do not usually claim that "the study's effect was significant," independent of any theoretical explanation. Instead, they tend to make more specific statistical inferences about (a) single tests of theory-based individual hypotheses and (b) multiple (union-intersection) tests of theory-based joint null hypotheses. *P*-hacking does not usually inflate the relevant Type I error rates in either of these cases.

### *P-Hacking May Increase Theoretical Errors, Not Statistical Errors*

To be clear, I am not denying that *p*-hacking occurs, and I am not arguing that it is always harmless. I am only arguing that it does not usually inflate relevant Type I error rates. In some cases, *p*-hacking may be problematic because it increases *theoretical errors*, rather than *statistical errors*, and it does so as a result of *biased selective reporting* (Rubin, 2020a, p. 383).

For example, in Simmons et al.'s (2011) scenario, the inference that listening to the song "When I'm Sixty-Four" makes people younger may represent a theoretical error rather than a statistical error. In particular, the *p*-hacked result may be a statistical *true positive* that has been misinterpreted as representing "chronological rejuvenation" when it actually represents Meehlian crud or some other statistically real but theoretically misleading effect (Meehl, 1990b, pp. 207-208). In this case, the problem is theoretical misinterpretation rather than Type I error rate inflation.

Theoretical errors are more likely to occur when the overall pattern of theoretically relevant evidence is obscured due to biased selective reporting. Hence, it is important to try to identify and reduce biased reporting through the use of open science practices such as open data, open research materials, and robustness, multiverse, and specification curve analyses. For example, reporting all of the evidence for and against chronological rejuvenation would be likely to show that the *p*-hacked result is part of a tiny minority of confirmatory evidence compared to a vast majority of disconfirmatory evidence, including other significant results that can be misinterpreted as showing the opposite of chronological rejuvenation (i.e., accelerated ageing!).

Importantly, theoretical errors can also be reduced through a rigorous critical evaluation of relevant theory. For example, in Simmons et al.'s (2011) scenario, we might ask: What larger theoretical framework explains "chronological rejuvenation"? What is the quality of that theory relative to other explanations for the results? How well does that theory justify the specific methodological and analytical decisions that the researcher made (e.g., including father's age as a covariate)? And what other evidence is there for and against the theory in the current study aside from a single ANCOVA result? These theoretical issues were not considered in Simmons et al.'s scenario. However, in practice, they would operate as an important (not infallible) line of defence against theoretical errors by helping to (a) screen out low-quality theories and (b) motivate and guide efforts to detect biased selective reporting (see also Simmons et al., 2011, p. 1363, Point 3; Rubin, 2017, p. 314; Syrjänen, 2023, p. 16).

Neyman and Pearson (1928, p. 232) cautioned that significance "tests should only be regarded as tools which must be used with discretion and understanding, and not as instruments which in themselves give the final verdict" (see also Bolles, 1962, p. 645; Boring, 1919, pp. 337-338; Chow, 1998, p. 169; Cox, 1958, p. 357; Hager, 2013, p. 261; Haig, 2018, p. 199; Lykken, 1968, p. 158; McShane et al., 2023; Meehl, 1978, p. 824; Meehl, 1997, p. 401; Szollosi & Donkin, 2021, p. 5). *P*-hacking is most problematic for those who ignore this advice and rely on *p* values as the sole arbiters of scientific decisions rather than as mere steppingstones on the way to making substantive theoretical inferences during a fallible process of inference to the best explanation (Haig, 2009; Mackonis, 2013).

## Optional Stopping

In the case of undisclosed optional stopping or data peeking, a researcher tests a hypothesis using a certain sample size and then collect more data and retest that same hypothesis using a larger sample size if their first test does not yield a significant result. They then continue this process until they obtain a significant result, at which point they report their significant result and hide their nonsignificant results.

Undisclosed optional stopping represents result-dependent multiple testing across a series of tests that have different sample sizes. A key concern here is that the $\alpha_{Constituent}$ for each test needs to be adjusted to account for the union-intersection testing of a joint null hypothesis that will be rejected when one of the tests yields a significant result. Failure to adjust $\alpha_{Constituent}$ will lead to inflation of the familywise error rate above $\alpha_{Joint}$. A further a concern is that, if the number of "stop-and-tests" is not specified in advance, then the actual familywise Type I error rate will be incalculable in repetitions of an exploratory optional stopping procedure. However, neither of these concerns is warranted.

First, a researcher who engages in undisclosed optional stopping has no choice but to limit their statistical inference to their final reported sample size because, by definition, they do not refer to any of their previous tests that used different sample sizes. In this case, it is appropriate for the researcher to use an unadjusted $\alpha_{Individual}$. Here, $\alpha_{Individual}$ refers to the frequency with which they would make an incorrect decision to reject the specified statistical null hypothesis during an imaginary long run of repeated random sampling in which samples are the same size as that used in the final reported test (e.g., $N$ = 300; Fraser, 2019, p. 140; Reid, 1995, p. 138). This long run would not include any of the other unreported tests that yielded nonsignificant results (e.g., *N*s = 270, 280, & 290) or any of the tests that might have occurred had the current test not yielded a significant result (e.g., *N*s = 310, 320, 330, etc.). Certainly, it is possible to make an inference about a joint null hypothesis that refers to other tests in the series (see below). However, this would represent a different statistical inference that is warranted by a different (familywise) Type I error rate. The current statistical inference is restricted to a reference set that excludes the unreported tests. As in the case of *p*-hacking, the fact that this inference is reported *because* it refers to a significant result does not alter the individual hypothetical probability of that result. To illustrate, imagine that you throw a 20-sided dice, hoping to get an "8," and you only throw the dice again if you fail to get an "8" on your previous throw. You finally get an "8" on your $20^{th}$ throw. In this case, it would be correct to report that you had a .050 probability of getting an "8" *on that particular throw* even if you did not report your first 19 unsuccessful throws. This individual (marginal) probability is not invalidated by the fact that the familywise (union) probability of getting an "8" in *at least one of the 20 throws* is .642. Furthermore, given that your probability statement refers to a single throw, a repetition of the associated procedure would only entail a single throw, and not 20 throws.

Second, if a researcher wanted to adjust their $\alpha_{Constituent}$ to maintain their familywise Type I error rate at $\alpha_{Joint}$ during the process of optional stopping, then they could do so without planning the number of stop-and-tests in advance. Again, relevant Type I error rates refer to reported statistical inferences and not to planned but unreported statistical inferences. Hence, for example, if a researcher planned to adjust their familywise Type I error rate for a series of five stop-and-tests but ended up deviating from that plan and making a statistical inference about a series of only three stop-and-tests, then they should adjust their $\alpha_{Constituent}$ based on $k = 3$, not $k = 5$. In this case, their actual familywise error rate for their specified ($k = 3$) statistical inference would match their $\alpha_{Joint}$ for that inference.

There may be a concern that the researcher may not report their "actual" number of stop-and-tests in the previous example. However, we should not be concerned about the overall number of tests that a researcher has happened to perform. Instead, we should be concerned about the "actual" number of tests that are formally associated with a specific statistical inference (i.e., $k$) and, in the current case, that number is three, not five.

Finally, as with *p*-hacking, there may also be a concern that a researcher's selective reporting is hiding relevant disconfirming evidence (i.e., biased selective reporting). It is debatable whether the null results from prior stop-and-tests represent "disconfirming evidence" given that null results represent the absence of evidence rather than evidence of absence (Altman & Bland, 1995). Nonetheless, even if null results are accepted as disconfirming evidence, the presence of this undisclosed evidence will not inflate the Type I error rate because, as discussed previously, hiding disconfirming evidence biases theoretical inferences, not statistical inferences. Furthermore, a significant result obtained via undisclosed optional stopping may either confirm or disconfirm a directional hypothesis. Hence, if a researcher stops data collection when they obtain a significant result, regardless of whether that result confirms or disconfirms their hypothesis, then their optional stopping will not bias their theoretical inference about their directional hypothesis (Rubin, 2020a, p. 381).

## Double Dipping

It has been proposed that the same data cannot be used to both generate and then test the same hypothesis (e.g., Nosek et al., 2018, p. 2600; Wagenmakers et al., 2012, p. 633). Engaging in this double dipping strategy is thought to inflate Type I error rates. For example, Wagenmakers et al. (2012, p. 633) argued that, "if you carry out a hypothesis test on the very data that inspired that test in the first place then the statistics are invalid… Whenever a researcher uses double-dipping strategies, Type I error rates will be inflated and *p* values can no longer be trusted."

Contrary to this argument, carrying out a hypothesis test on the same data that inspired the test does not necessarily invalidate the statistics. For example, it is perfectly acceptable to use the result from one statistical test to create a statistical null hypothesis for a second test which is then tested using the same data as long as the second test's result is independent from the first test's result. The logical problem of circularity only occurs when the same result is used to both (a) support the theoretical rationale for a hypothesis and (b) claim additional support for that hypothesis (Devezer et al., 2021; Kriegeskorte et al., 2009, p. 535; Rubin & Donkin, 2022, pp. 5-6; Spanos, 2010, p. 216; Worrall, 2010, p. 131). Furthermore, this problem of circularity represents a theoretical error, rather than a statistical error. Consequently, although double dipping may sometimes invalidate theoretical inferences, it does not inflate Type I error rates.

## HARKing

Hypothesizing after the results are known, or HARKing, refers to the questionable research practice of "presenting post hoc hypotheses in a research report as if they were, in fact, a priori hypotheses" (Kerr, 1998, p. 197). HARKing is thought to inflate Type I error rates (e.g., Bergkvist, 2020; Stefan & Schönbrodt, 2023, p. 4). However, HARKing represents post hoc *theorizing*, and so it affects theoretical inferences rather than statistical inferences. Indeed, in his seminal article on the subject, Kerr (1998, p. 205) did not argue that HARKing inflates Type I error rates. Instead, his concern was that, when "a Type I error is followed by HARKing, then 'theory' is constructed to account for what is, in fact, an illusory effect" (p. 205). In other words, he was not concerned that HARKing inflates Type I error rates, but rather that it may be used to "translate Type I errors into theory" (Kerr, 1998, p. 205).

Kerr (1998) conceded that Type I errors can also be translated into theory following explicit, transparent, post hoc theorizing, rather than undisclosed HARKing. However, he believed that the translation is more problematic following HARKing because "an explicitly post hoc hypothesis implicitly acknowledges its dependence upon the result in hand as a cornerstone (or perhaps, the entirety) of its foundation, and thereby sensitizes the reader to the vulnerability of the hypothesis to the risks of an immediate Type I error" (Kerr, 1998, p. 205). Contrary to this reasoning, "the risks of an immediate Type I error" do not vary as a function of either the origin or quality of a hypothesis. To believe that they do is to commit the Bayesian inversion fallacy. Hence, there is no reason to believe that HARKing either inflates Type I error rates or that it exacerbates the costs of Type I errors.

# Evidence for Type I Error Rate Inflation?

What about the evidence of Type I error rate inflation? There are two problems with this evidence that threaten its validity.

First, similar to Simmons et al.'s (2011) demonstration, evidence of Type I error rate inflation tends to confound statistical inferences about *individual* null hypotheses with statistical inferences about *joint* null hypotheses. For example, simulations of actual Type I error rates compute the familywise error rate for a joint null hypothesis and then apply that error rate to individual null hypotheses, claiming that, because the familywise error rate is, for example, .143, there is a .143 chance of incorrectly rejecting *each individual* null hypothesis. Again, this reasoning is widely acknowledged to be incorrect (Armstrong, 2014, p. 505; Cook & Farewell, 1996, pp. 96–97; Fisher, 1971, p. 206; García-Pérez, 2023, p. 15; Greenland, 2021, p. 5; Hewes, 2003, p. 450; Hurlbert & Lombardi, 2012, p. 30; Matsunaga, 2007, p. 255; Molloy et al., 2022, p. 2; Parker & Weir, 2020, p. 564; Parker & Weir, 2022, p. 2; Rothman, 1990, p. 45; Rubin, 2017b, pp. 271–272; Rubin, 2020a, p. 380; Rubin, 2021a, 2021c, pp. 10978-10983; Rubin, 2024, p. 3; Savitz & Olshan, 1995, p. 906; Senn, 2007, pp. 150-151; Sinclair et al., 2013, p. 19; Tukey, 1953, p. 82; Turkheimer et al., 2004, p. 727; Veazie, 2006, p. 809; Wilson, 1962, p. 299).

Second, evidence of Type I error rate inflation may also depend on a fallacious comparison between (a) the probability of rejecting a null hypothesis when it is true and (b) the probability of a null hypothesis being true when it is rejected (Pollard & Richardson, 1987). As discussed previously, the first probability is equivalent to a frequentist Type I error rate: Pr(reject $H_0$ ; $H_0$ is true). However, the second probability does not provide an appropriate benchmark against which to judge Type I error rate inflation because it represents a conditional posterior probability about the truth of the null hypothesis: Pr($H_0$ is true | reject $H_0$). Hence, showing that Pr($H_0$ is true | reject $H_0$) > Pr(reject $H_0$ ; $H_0$ is true) does not provide a valid demonstration of Type I error rate inflation. Instead, it demonstrates the Bayesian inversion fallacy because it confuses the unconditional probability of rejecting a true null hypothesis with the conditional probability that a null hypothesis is true given that it has been rejected (Gigerenzer, 2018; Greenland et al., 2016; Mayo & Morey, 2017; Pollard & Richardson, 1987).

# Implications for the Replication Crisis

Type I error rate inflation may not be a major contributor to the replication crisis. Certainly, some failed replications may be due to Type I errors in original studies. However, actual Type I error rates are rarely inflated above their nominal levels, and so the level of Type I errors in a field is liable to be around that field's conventional nominal level (see also Neyman, 1977, p. 108). Hence, Type I error rate inflation cannot explain unexpectedly low replication rates.

In contrast, theoretical errors may be higher than expected. In particular, unacknowledged misinterpretations of theory, methodology, data, and analyses may all inflate theoretical errors above their "nominal" expected level, resulting in incorrect theoretical inferences and unexpectedly low replication rates. For example, researchers may assume a higher degree of theoretical equivalence between an original study and a "direct" replication than is warranted. A failed replication may then represent the influence of an unrecognized "hidden moderator" that produces a true positive result in the original study and a true negative result in the replication study. Of course, scientists should attempt to specify and investigate such hidden moderators in future studies (Klein et al., 2018, p. 482). Nonetheless, ignoring hidden moderators does not mitigate their deleterious impact on replicability!

From this perspective, the replication crisis may be explained, at least in part, by researchers' underestimation of theoretical errors and their misinterpretation of statistical errors (i.e., statistical reification; Greenland, 2017b, 2023; see also Brower, 1949; Gigerenzer, 1993). Specifically, researchers may overestimate (a) their theoretical understanding of effects and (b) the extent to which a Type I error rate implies replicability. These two issues may combine to produce overconfident researchers who have unrealistically high expectations about replication rates during "direct" replications (Rubin, 2021, pp. 5828-5829). Accordingly, an appropriate response to the replication crisis is for researchers to adopt a more modest perspective that recognizes (a) the important role of scientific ignorance during theoretical inferences (e.g., Feynman, 1955; Firestein, 2012; Merton, 1987) and (b) the limited scope of Type I error rates during statistical inferences (e.g., Bolles, 1962; Cox, 1958; Fisher, 1926; Greenland, 2017a, 2017b). This more modest perspective may help to provide more realistic expectations about replication rates and a better appreciation of replication failures as a vital aspect of scientific progress (Barrett, 2015; Redish et al., 2018; Rubin, 2021b).

# Summary and Conclusion

The replication crisis has been partly explained in terms of Type I error rate inflation. In particular, it has been argued that questionable research practices inflate actual Type I error rates above their nominal levels, leading to an unexpectedly high level of false positives in the literature and, consequently, unexpectedly low replication rates. In this article, I have offered the alternative view that questionable and other research practices do not usually inflate relevant Type I error rates.

During significance testing, each statistical inference is assigned a nominal Type I error rate or alpha level. Type I error rate inflation occurs if the *actual* Type I error rate for that inference is higher than the *nominal* error rate. The actual Type I error rate can be calculated using the formula $1 - (1 - \alpha)^k$, in which $k$ is the number of significance tests that are used to make the statistical inference. I have argued that the actual Type I error rate is not usually inflated above the nominal rate and that, when it is, the inflation is transparent and easily resolved because $k$ is known by readers. Indeed, $k$ must be known by readers because the researcher must formally associate their statistical inference with one or more significance tests, and $k$ is the number of those tests. A key point here is that $k$ is not the number of tests that a researcher conducted, including those that they conducted and did not report. Instead, $k$ is the number of tests that the researcher formally associates with a statistical inference about a specified null hypothesis.

It is true that some actual Type I error rates may be above a field's conventional alpha level. However, this issue does not necessarily represent Type I error rate inflation. Type I error rate inflation only occurs when the actual Type I error rate for a specified statistical inference is higher than the nominal Type I error rate for that inference, regardless of whether that nominal rate is higher or lower than the conventional level.

It is true that the Type I error rate for a researcher's specified statistical inference about a particular individual or joint null hypothesis may be different to the Type I error rates for other statistical inferences that they could, would, or should have made about other individual or joint null hypotheses or for other statistical inferences that they planned to make or actually made and then either reported or failed to report. It is also true that different researchers may disagree about which are the most appropriate or theoretically relevant statistical inferences or alpha levels in any given research situation. However, none of these points imply that the actual Type I error rate for a researcher's reported statistical inference is inflated above the alpha level that they have set for that particular inference.

It is true that the actual familywise Type I error rate is always above $\alpha_{Constituent}$. However, researchers can adjust $\alpha_{Constituent}$ to avoid Type I error rate inflation with respect to their statistical inferences about associated joint null and, if they do not adjust $\alpha_{Constituent}$, then the extent of the inflation will be transparent to others and easily resolved. Hence, this potential issue is not problematic.

It is also true that the *studywise* Type I error rate is always above $\alpha_{Constituent}$. However, researchers do not usually make statistical inferences about joint studywise null hypotheses, and so this point is usually irrelevant. Nonetheless, if a studywise error rate does become relevant, then it can be identified and controlled by adjusting $\alpha_{Constituent}$. This adjustment is even possible if researchers take the unusual step of making relatively vague and atheoretical statistical inferences about exploratory joint studywise null hypotheses (e.g., "the study's effect was significant") because, even in this case, they will need to specify the relevant statistical tests that they are using to make this inference.

Finally, it is true that a researcher's probability of incorrectly rejecting a substantive null hypothesis and incorrectly accepting a substantive alternative hypothesis may be greater than their alpha level because this probability may be influenced by theoretical errors as well as Type I errors. However, these theoretical errors cannot be said to inflate Type I error rates because Type I error rates refer to random sampling error per se. They do not account for theoretical errors.

Based on these points, I have argued that the following research practices do not usually inflate relevant Type I error rates: model misspecification, multiple testing, selective inference, forking paths, exploratory analyses, *p*-hacking, optional stopping, double dipping, and HARKing. My view is consistent with a logico-historical approach to hypothesis testing that considers the logical relations between a reported hypothesis and its test results independent from the psychological origin of the hypothesis and results (e.g., researcher bias; Musgrave, 1974, p. 17; Reichenbach, 1938, p. 5; Popper, 1962, p. 140; Popper, 2002 p. 7). Hence, contrary to Mayo's (1996, 2018) error statistical approach, my logical inference-based approach denies that "biasing selection effects" in the testing context inflate Type I error rates. Instead, it offers a more restricted conceptualization of Type I error rates as indicators of the frequency of incorrect decisions in hypothetical and potentially unplanned testing situations in which random sampling error is the only source of decision-making error. This more modest conceptualization may help to provide more realistic expectations about replication rates and, consequently, less concern about replication failures.

## Endnotes

1. The semicolon in "Pr(reject $H_0$ **;** $H_0$ is true)" is used to indicate that "$H_0$ is true" is a fixed assumption, rather than a random variable that can be true or false. Hence, Pr(reject $H_0$ **;** $H_0$ is true) is an unconditional probability rather than a conditional probability. In contrast, the vertical bar in "Pr($H_0$ is true **|** reject $H_0$)" is used to indicate a conditional probability (Mayo & Morey, 2017, Footnote 2; Mayo & Spanos, 2006, p. 331; Wasserman, 2013).

2. Fisher (1930, p. 530) explained that the Bayesian approach of "inverse probability" is applicable when "we know that the population from which our observations were drawn had itself been drawn at random from a super-population of known specification" (e.g., a superpopulation of 200 null populations of which 100 are known to be true and 100 are known to be false). Hence, as Cox (1958) explained, "if the population sampled has itself been selected by a random procedure with known prior probabilities, it seems to be generally agreed that inference should be made using Bayes's theorem" (pp. 357-358).

# References

Altman, D. G., & Bland, J. M. (1995). Absence of evidence is not evidence of absence. *British Medical Journal*, *311*(7003), 485–485. https://doi.org/10.1136/bmj.311.7003.485

Armstrong, R. A. (2014). When to use the Bonferroni correction. *Ophthalmic and Physiological Optics*, *34*(5), 502-508. https://doi.org/10.1111/opo.12131

Barrett, L. F. (2015). Psychology is not in crisis. *The New York Times*. https://www.nytimes.com/2015/09/01/opinion/psychology-is-not-in-crisis.html

Bender, R., & Lange, S. (2001). Adjusting for multiple testing—when and how? *Journal of Clinical Epidemiology*, *54*(4), 343-349. https://doi.org/10.1016/S0895-4356(00)00314-0

Benjamin, D. J., Berger, J. O., Johannesson, M., Nosek, B. A., Wagenmakers, E. J., Berk, R., Bollen, K. A., Brembs, B., Brown, L., Camerer, C., Cesarini, D., Chambers, C. D., Clyde, M., Cook, T. D., De Boeck, P., Dienes, Z., Dreber, A., Easwaran, K., Efferson, C., ... & Johnson, V. E. (2018). Redefine statistical significance. *Nature Human Behaviour*, *2*(1), 6-10. https://doi.org/10.1038/s41562-017-0189-z

Bergkvist, L. (2020). Preregistration as a way to limit questionable research practice in advertising research. *International Journal of Advertising*, *39*(7), 1172-1180. https://doi.org/10.1080/02650487.2020.1753441

Berk, R. A., Western, B., & Weiss, R. E. (1995). Statistical inference for apparent populations. *Sociological Methodology, 25*, 421-458. https://doi.org/10.2307/271073

Birnbaum, A. (1962). On the foundations of statistical inference. *Journal of the American Statistical Association*, *57*(298), 269-306. https://doi.org/10.1080/01621459.1962.10480660

Bolles, R. C. (1962). The difference between statistical hypotheses and scientific hypotheses. *Psychological Reports*, *11*(3), 639-645. https://doi.org/10.2466/pr0.1962.11.3.639

Boring, E. G. (1919). Mathematical vs. scientific significance. *Psychological Bulletin*, *16*(10), 335-338. https://doi.org/10.1037/h0074554

Brower, D. (1949). The problem of quantification in psychological science. *Psychological Review*, *56*(6), 325–333. https://doi.org/10.1037/h0061802

Chow, S. L. (1998). Précis of statistical significance: Rationale, validity, and utility. *Behavioral and Brain Sciences*, *21*(2), 169-194. https://doi.org/10.1017/S0140525X98001162

Cook, R. J., & Farewell, V. T. (1996). Multiplicity considerations in the design and analysis of clinical trials. *Journal of the Royal Statistical Society: Series A (Statistics in Society)*, *159*(1), 93-110. http://doi.org/10.2307/2983471

Cox, D. R. (1958). Some problems connected with statistical inference. *Annals of Mathematical Statistics*, *29*(2), 357-372. http://doi.org/10.1214/aoms/1177706618

Cox, D. R., & Mayo, D. G. (2010). Objectivity and conditionality in frequentist inference. In D. G. Mayo & A. Spanos (Eds.), *Error and inference: Recent exchanges on experimental* *reasoning, reliability, and the objectivity and rationality of science* (pp. 276-304). Cambridge University Press. http://doi.org/10.1017/CBO9780511657528

Del Giudice, M., & Gangestad, S. W. (2021). A traveler's guide to the multiverse: Promises, pitfalls, and a framework for the evaluation of analytic decisions. *Advances in Methods and Practices in Psychological Science, 4*(1). https://doi.org/10.1177/2515245920954925

Dennis, B., Ponciano, J. M., Taper, M. L., & Lele, S. R. (2019). Errors in statistical inference under model misspecification: Evidence, hypothesis testing, and AIC. *Frontiers in Ecology and Evolution*, *7*, Article 372. https://doi.org/10.3389/fevo.2019.00372

Devezer, B., & Buzbas, E. O. (2023). Rigorous exploration in a model-centric science via epistemic iteration. *Journal of Applied Research in Memory and Cognition*, *12*(2), 189–194. https://doi.org/10.1037/mac0000121

Devezer, B., Navarro, D. J., Vandekerckhove, J., & Buzbas, E. O. (2021). The case for formal methodology in scientific reform. *Royal Society Open Science, 8*(3), Article 200805. https://doi.org/10.1098/rsos.200805

Feynman, R. P. (1955). The value of science. *Engineering and Science*, *19*(3), 13-15. https://calteches.library.caltech.edu/1575/1/Science.pdf

Firestein, S. (2012). *Ignorance: How it drives science*. Oxford University Press.

Fisher, R. A. (1922). On the mathematical foundations of theoretical statistics. *Philosophical Transactions of the Royal Society of London. Series A, Containing Papers of a Mathematical or Physical Character*, *222*, 309-368. https://doi.org/10.1098/rsta.1922.0009

Fisher, R. A. (1926). The arrangement of field experiments. *Journal of the Ministry of Agriculture*. *33*, 503-515. https://doi.org/10.23637/rothamsted.8v61q

Fisher, R. A. (1930). Inverse probability. *Mathematical Proceedings of the Cambridge Philosophical Society*, *26*(4), 528-535. https://doi.org/10.1017/S0305004100016297

Fisher, R. A. (1956). *Statistical methods and scientific inference*. Oliver & Boyd.

Fisher, R. A. (1971). *The design of experiments* (9th ed.). Hafner Press.

Fraser, D. A. S. (2019). The *p*-value function and statistical inference. *The American Statistician*, *73*(sup1), 135-147. https://doi.org/10.1080/00031305.2018.1556735

García-Pérez, M. A. (2023). Use and misuse of corrections for multiple testing. *Methods in Psychology*, *8*, Article 100120. https://doi.org/10.1016/j.metip.2023.100120

Gelman, A., & Loken, E. (2014). The statistical crisis in science. *American Scientist*, *102*, Article 460. http://dx.doi.org/10.1511/2014.111.460

Gigerenzer, G. (1993). The superego, the ego, and the id in statistical reasoning. In G. Keren & C. Lewis (Eds.), *A handbook for data analysis in the behavioral sciences: Methodological issues* (pp. 311–339). Lawrence Erlbaum Associates.

Gigerenzer, G. (2018). Statistical rituals: The replication delusion and how we got there. *Advances in Methods and Practices in Psychological Science*, *1*(2), 198-218. https://doi.org/10.1177/2515245918771329

Greenland, S. (2017a). For and against methodologies: Some perspectives on recent causal and statistical inference debates. *European Journal of Epidemiology*, *32*, 3-20. https://doi.org/10.1007/s10654-017-0230-6

Greenland, S. (2017b). Invited commentary: The need for cognitive science in methodology. *American Journal of Epidemiology*, *186*(6), 639-645. https://doi.org/10.1093/aje/kwx259

Greenland, S. (2021). Analysis goals, error-cost sensitivity, and analysis hacking: Essential considerations in hypothesis testing and multiple comparisons. *Paediatric and Perinatal Epidemiology, 35*(1), 8-23. https://doi.org/10.1111/ppe.12711

Greenland, S. (2023). Connecting simple and precise *p*-values to complex and ambiguous realities. *Scandinavian Journal of Statistics, 50*(3), 899-914. https://doi.org/10.1111/sjos.12645

Greenland, S., Senn, S. J., Rothman, K. J., Carlin, J. B., Poole, C., Goodman, S. N., & Altman, D. G. (2016). Statistical tests, *P* values, confidence intervals, and power: A guide to misinterpretations. *European Journal of Epidemiology*, *31*, 337-350. https://doi.org/10.1007/s10654-016-0149-3

Hager, W. (2013). The statistical theories of Fisher and of Neyman and Pearson: A methodological perspective. *Theory & Psychology, 23*(2), 251-270. https://doi.org/10.1177/0959354312465483

Haig, B. D. (2009). Inference to the best explanation: A neglected approach to theory appraisal in psychology. *The American Journal of Psychology*, *122*(2), 219–234. https://doi.org/10.2307/27784393

Haig, B. D. (2018). *Method matters in psychology: Essays in applied philosophy of science*. Springer. https://doi.org/10.1007/978-3-030-01051-5

Hancock, G. R., & Klockars, A. J. (1996). The quest for α: Developments in multiple comparison procedures in the quarter century since. *Review of Educational Research, 66*(3), 269-306. https://doi.org/10.3102/00346543066003269

Hewes, D. E. (2003). Methods as tools: A response to O'Keefe. *Human Communication Research, 29*(3), 448-454. https://doi.org/10.1111/j.1468-2958.2003.tb00847.x

Hitchcock, C., & Sober, E. (2004). Prediction versus accommodation and the risk of overfitting. *British Journal for the Philosophy of Science, 55*(1), 1-34. https://doi.org/10.1093/bjps/55.1.1

Hochberg, Y., & Tamrane, A. C. (1987). *Multiple comparison procedures.* Wiley. http://doi.org/10.1002/9780470316672

Hurlbert, S. H., & Lombardi, C. M. (2012). Lopsided reasoning on lopsided tests and multiple comparisons. *Australian & New Zealand Journal of Statistics*, *54*(1), 23-42. https://doi.org/10.1111/j.1467-842X.2012.00652.x

John, L. K., Loewenstein, G., & Prelec, D. (2012). Measuring the prevalence of questionable research practices with incentives for truth telling. *Psychological Science, 23*(5), 524-532. https://doi.org/10.1177/0956797611430953

Kerr, N. L. (1998). HARKing: Hypothesizing after the results are known. *Personality and Social Psychology Review, 2*(3), 196-217. http://doi.org/10.1207/s15327957pspr0203_4

Kim, K., Zakharkin, S. O., Loraine, A., & Allison, D. B. (2004). Picking the most likely candidates for further development: Novel intersection-union tests for addressing multi-component hypotheses in comparative genomics. *Proceedings of the American Statistical Association, ASA Section on ENAR Spring Meeting* (pp. 1396-1402). http://www.uab.edu/cngi/pdf/2004/JSM%202004%20-IUTs%20Kim%20et%20al.pdf

Klein, R. A., Vianello, M., Hasselman, F., Adams, B. G., Adams Jr, R. B., Alper, S., Aveyard, M., Axt, J. R., Babalola, M. T., Bahník, S., Batra, R., Berkics, M., Bernstein, M. J., Berry, D. R., Bialobrzeska, O., Binan, E. D., Bocian, K., Brandt, M. J., Busching, R., ... & Sowden, W. (2018). Many Labs 2: Investigating variation in replicability across samples and settings. *Advances in Methods and Practices in Psychological Science, 1*(4), 443-490. https://doi.org/10.1177/2515245918810225

Kotzen, M. (2013). Multiple studies and evidential defeat. *Noûs, 47*(1), 154-180. http://www.jstor.org/stable/43828821
Kriegeskorte, N., Simmons, W. K., Bellgowan, P. S., & Baker, C. I. (2009). Circular analysis in systems neuroscience: The dangers of double dipping. *Nature Neuroscience*, *12*(5), 535-540. https://doi.org/10.1038/nn.2303
Kuhn, T. S. (1977). *The essential tension: Selected studies in the scientific tradition and change.* The University of Chicago Press.
Lehmann, E. L. (1993). The Fisher, Neyman-Pearson theories of testing hypotheses: One theory or two? *Journal of the American Statistical Association*, *88*(424), 1242-1249. https://doi.org/10.1080/01621459.1993.10476404
Lykken, D. T. (1968). Statistical significance in psychological research. *Psychological Bulletin*, *70*(3), 151-159. https://doi.org/10.1037/h0026141
Mackonis, A. (2013). Inference to the best explanation, coherence and other explanatory virtues. *Synthese, 190*(6), 975-995. https://doi.org/10.1007/s11229-011-0054-y
Maier, M., & Lakens, D. (2022). Justify your alpha: A primer on two practical approaches. *Advances in Methods and Practices in Psychological Science*, *5*(2). https://doi.org/10.1177/25152459221080396
Matsunaga, M. (2007). Familywise error in multiple comparisons: Disentangling a knot through a critique of O'Keefe's arguments against alpha adjustment. *Communication Methods and Measures*, *1*(4), 243-265. https://doi.org/10.1080/19312450701641409
Mayo, D. G. (1996). *Error and the growth of experimental knowledge*. The University of Chicago Press.
Mayo, D. G. (2014). On the Birnbaum argument for the strong likelihood principle. *Statistical Science*, *29*, 227-239. http://doi.org/10.1214/13-STS457
Mayo, D. G. (2018). *Statistical inference as severe testing: How to get beyond the statistics wars*. Cambridge University Press. http://doi.org/10.1017/9781107286184
Mayo, D. G., & Morey, R. D. (2017). *A poor prognosis for the diagnostic screening critique of statistical tests.* OSFPreprints. https://doi.org/10.31219/osf.io/ps38b
Mayo, D. G., & Spanos, A. (2006). Severe testing as a basic concept in a Neyman–Pearson philosophy of induction. *The British Journal for the Philosophy of Science*, *57*(2), 323-357. https://doi.org/10.1093/bjps/axl003
McShane, B. B., Bradlow, E. T., Lynch, J. G. Jr., & Meyer, R. J. (2023). "Statistical significance" and statistical reporting: Moving beyond binary. *Journal of Marketing*, *88*(3), 1-19. http://doi.org/10.1177/00222429231216910
Meehl, P. E. (1978). Theoretical risks and tabular asterisks: Sir Karl, Sir Ronald, and the slow progress of soft psychology. *Journal of Consulting and Clinical Psychology*, *46*(4), 806-834. https://doi.org/10.1037/0022-006X.46.4.806
Meehl, P. E. (1990a). Appraising and amending theories: The strategy of Lakatosian defense and two principles that warrant it. *Psychological Inquiry*, *1*(2), 108-141. http://www.jstor.org/stable/1448768
Meehl, P. E. (1990b). Why summaries of research on psychological theories are often uninterpretable. *Psychological Reports*, *66*(1), 195–244. https://doi.org/10.2466/pr0.1990.66.1.195
Meehl, P. E. (1997). The problem is epistemology, not statistics: Replace significance tests by confidence intervals and quantify accuracy of risky numerical predictions. In L. L. Harlow,

S. A. Mulaik, & J. H. Steiger (Eds.), *What if there were no significance tests?* (pp. 393–425). Lawrence Erlbaum Associates.
Merton, R. K. (1987). Three fragments from a sociologist's notebooks: Establishing the phenomenon, specified ignorance, and strategic research materials. *Annual Review of Sociology, 13*(1), 1-29. https://doi.org/10.1146/annurev.so.13.080187.000245
Molloy, S. F., White, I. R., Nunn, A. J., Hayes, R., Wang, D., & Harrison, T. S. (2022). Multiplicity adjustments in parallel-group multi-arm trials sharing a control group: Clear guidance is needed. *Contemporary Clinical Trials*, *113*, Article 106656. https://doi.org/10.1016/j.cct.2021.106656
Morgan, J. F. (2007). *P* value fetishism and use of the Bonferroni adjustment. *Evidence-Based Mental Health*, *10*, 34-35. http://doi.org/10.1136/ebmh.10.2.34
Munafò, M. R., Nosek, B. A., Bishop, D. V., Button, K. S., Chambers, C. D., Percie du Sert, N., Simonsohn, U., Wagenmakers, E. J., Ware, J. J., & Ioannidis, J. (2017). A manifesto for reproducible science. *Nature Human Behaviour*, *1*, Article 0021. https://doi.org/10.1038/s41562-016-0021
Musgrave, A. (1974). Logical versus historical theories of confirmation. *The British Journal for the Philosophy of Science*, *25*(1), 1-23. https://doi.org/10.1093/bjps/25.1.1
Neyman, J. (1950). *First course in probability and statistics.* Henry Holt.
Neyman, J. (1977). Frequentist probability and frequentist statistics. *Synthese, 36*, 97–131. https://doi.org/10.1007/BF00485695
Neyman, J., & Pearson, E. S. (1928). On the use and interpretation of certain test criteria for purposes of statistical inference: Part I. *Biometrika 20A*(1/2), 175–240. http://doi.org/10.2307/2331945
Neyman, J., & Pearson, E. S. (1933). IX. On the problem of the most efficient tests of statistical hypotheses. *Philosophical Transactions of the Royal Society A, 231*, 289-337. https://doi.org/10.1098/rsta.1933.0009
Nosek, B. A., Ebersole, C. R., DeHaven, A. C., & Mellor, D. T. (2018). The preregistration revolution. *Proceedings of the National Academy of Sciences*, *115*(11), 2600-2606. https://doi.org/10.1073/pnas.1708274114
Nosek, B. A., Hardwicke, T. E., Moshontz, H., Allard, A., Corker, K. S., Dreber, A., Fidler, F., Hilgard, J., Kline Struhl, M., Nuijten, M. B., Rohrer, J. M., Romero, F., Scheel, A. M., Scherer, L. D., Schönbrodt, F. D., & Vazire, S. (2022). Replicability, robustness, and reproducibility in psychological science. *Annual Review of Psychology, 73*, 719-748. https://doi.org/10.1146/annurev-psych-020821-114157
Nosek, B. A., & Lakens, D. (2014). Registered reports. *Social Psychology*, *45*(3), 137-141. https://doi.org/10.1027/1864-9335/a000192
Oberauer, K., & Lewandowsky, S. (2019). Addressing the theory crisis in psychology. *Psychonomic Bulletin & Review, 26*(5), 1596-1618. https://doi.org/10.3758/s13423-019-01645-2
Open Science Collaboration. (2015). Estimating the reproducibility of psychological science. *Science, 349*(6251), Article aac4716. https://doi.org/10.1126/science.aac4716
Parker, R. A., & Weir, C. J. (2020). Non-adjustment for multiple testing in multi-arm trials of distinct treatments: Rationale and justification. *Clinical Trials*, *17*(5), 562-566. https://doi.org/10.1177/1740774520941419
Parker, R. A., & Weir, C. J. (2022). Multiple secondary outcome analyses: Precise interpretation is important. *Trials*, *23*, Article 27. https://doi.org/10.1186/s13063-021-05975-2

Parker, T. H., Forstmeier, W., Koricheva, J., Fidler, F., Hadfield, J. D., Chee, Y. E., Kelly, C. D., Gurevitch, J., & Nakagawa, S. (2016). Transparency in ecology and evolution: Real problems, real solutions. *Trends in Ecology & Evolution*, *31*(9), 711-719. https://doi.org/10.1016/j.tree.2016.07.002

Perneger, T. V. (1998). What's wrong with Bonferroni adjustments. *British Medical Journal, 316*, 1236-1238. https://doi.org/10.1136/bmj.316.7139.1236

Pollard, P., & Richardson, J. T. (1987). On the probability of making Type I errors. *Psychological Bulletin, 102*(1), 159-163. https://doi.org/10.1037/0033-2909.102.1.159

Popper, K. R. (1962). *Conjectures and refutations: The growth of scientific knowledge.* Basic Books.

Popper, K. R. (2002). *The logic of scientific discovery.* Routledge.

Redish, D. A., Kummerfeld, E., Morris, R. L., & Love, A. C. (2018). Reproducibility failures are essential to scientific inquiry. *Proceedings of the National Academy of Sciences, 115*(20), 5042–5046. https://doi.org/10.1073/pnas.1806370115

Reichenbach, H. (1938). *Experience and prediction: An analysis of the foundations and the structure of knowledge.* The University of Chicago Press. https://philarchive.org/archive/REIEAP-2

Reid, N. (1995). The roles of conditioning in inference. *Statistical Science*, *10*(2), 138-157. https://doi.org/10.1214/ss/1177010027

Reid, N., & Cox, D. R. (2015). On some principles of statistical inference. *International Statistical Review*, *83*(2), 293-308. http:// doi.org/10.1111/insr.12067

Rothman, K. J. (1990). No adjustments are needed for multiple comparisons. *Epidemiology, 1*(1), 43-46. https://www.jstor.org/stable/20065622

Rothman, K. J., Greenland, S., & Lash, T. L. (2008). *Modern epidemiology* (3rd ed.). Lippincott Williams & Wilkins.

Rubin, M. (2017a). An evaluation of four solutions to the forking paths problem: Adjusted alpha, preregistration, sensitivity analyses, and abandoning the Neyman-Pearson approach. *Review of General Psychology, 21*(4), 321-329. https://doi.org/10.1037/gpr0000135

Rubin, M. (2017b). Do *p* values lose their meaning in exploratory analyses? It depends how you define the familywise error rate. *Review of General Psychology*, *21*(3), 269-275. https://doi.org/10.1037/gpr0000123

Rubin, M. (2020a). Does preregistration improve the credibility of research findings? *The Quantitative Methods for Psychology*, *16*(4), 376–390. https://doi.org/10.20982/tqmp.16.4.p376

Rubin, M. (2020b). "Repeated sampling from the same population?" A critique of Neyman and Pearson's responses to Fisher. *European Journal for Philosophy of Science*, *10,* Article 42, 1-15. https://doi.org/10.1007/s13194-020-00309-6

Rubin, M. (2021a). There's no need to lower the significance threshold when conducting single tests of multiple individual hypotheses. *Academia Letters*, Article 610. https://doi.org/10.20935/AL610

Rubin, M. (2021b). What type of Type I error? Contrasting the Neyman-Pearson and Fisherian approaches in the context of exact and direct replications. *Synthese*, *198*, 5809–5834. https://doi.org/10.1007/s11229-019-02433-0

Rubin, M. (2021c). When to adjust alpha during multiple testing: A consideration of disjunction, conjunction, and individual testing. *Synthese*, *199*, 10969–11000. https://doi.org/10.1007/s11229-021-03276-4

Rubin, M. (2022). The costs of HARKing. *British Journal for the Philosophy of Science, 73*(2), 535-560. https://doi.org/10.1093/bjps/axz050

Rubin, M. (2024). Inconsistent multiple testing corrections: The fallacy of using family-based error rates to make inferences about individual hypotheses. *Methods in Psychology, 10,* Article 100140. https://doi.org/10.1016/j.metip.2024.100140

Rubin, M., & Donkin, C. (2022). Exploratory hypothesis tests can be more compelling than confirmatory hypothesis tests. *Philosophical Psychology,* 38(7), 2019-2047. https://doi.org/10.1080/09515089.2022.2113771

Savitz, D. A., & Olshan, A. F. (1995). Multiple comparisons and related issues in the interpretation of epidemiologic data. *American Journal of Epidemiology, 142*(9), 904-908. https://doi.org/10.1093/oxfordjournals.aje.a117737

Schulz, K. F., & Grimes, D. A. (2005). Multiplicity in randomised trials I: Endpoints and treatments. *The Lancet, 365*(9470), 1591-1595. https://doi.org/10.1016/S0140-6736(05)66461-6

Senn, S. (2007). *Statistical issues in drug development* (2nd ed.). Wiley.

Simmons, J. P., Nelson, L. D., & Simonsohn, U. (2011). False-positive psychology: Undisclosed flexibility in data collection and analysis allows presenting anything as significant. *Psychological Science, 22*(11), 1359-1366. https://doi.org/10.1177/0956797611417632

Simonsohn, U., Simmons, J. P., & Nelson, L. D. (2020). Specification curve analysis. *Nature Human Behaviour*, *4*(11), 1208-1214. https://doi.org/10.1038/s41562-020-0912-z

Sinclair, J., Taylor, P. J., & Hobbs, S. J. (2013). Alpha level adjustments for multiple dependent variable analyses and their applicability—A review. *International Journal of Sports Science Engineering*, *7*(1), 17-20.

Spanos, A. (2006). Where do statistical models come from? Revisiting the problem of specification. *Optimality, 49,* 98-119. https://doi.org/10.1214/074921706000000419

Spanos, A. (2010). Akaike-type criteria and the reliability of inference: Model selection versus statistical model specification. *Journal of Econometrics*, *158*(2), 204-220. https://doi.org/10.1016/j.jeconom.2010.01.011

Steegen, S., Tuerlinckx, F., Gelman, A., & Vanpaemel, W. (2016). Increasing transparency through a multiverse analysis. *Perspectives on Psychological Science, 11*(5), 702-712. https://doi.org/10.1177/1745691616658637

Stefan, A. M., & Schönbrodt, F. D. (2023). Big little lies: A compendium and simulation of p-hacking strategies. *Royal Society Open Science*, *10*(2), Article 220346. https://doi.org/10.1098/rsos.220346

Syrjänen, P. (2023). Novel prediction and the problem of low-quality accommodation. *Synthese*, 202, Article 182. https://doi.org/10.1007/s11229-023-04400-2

Szollosi, A., & Donkin, C. (2021). Arrested theory development: The misguided distinction between exploratory and confirmatory research. *Perspectives on Psychological Science, 16*(4), 717-724. https://doi.org/10.1177/1745691620966796

Taylor, J., & Tibshirani, R. J. (2015). Statistical learning and selective inference. *Proceedings of the National Academy of Sciences*, *112*(25), 7629-7634. https://doi.org/10.1073/pnas.1507583112

Tukey, J. W. (1953). *The problem of multiple comparisons.* Princeton University.

Turkheimer, F. E., Aston, J. A., & Cunningham, V. J. (2004). On the logic of hypothesis testing in functional imaging. *European Journal of Nuclear Medicine and Molecular Imaging, 31*, 725-732. https://doi.org/10.1007/s00259-003-1387-7

Uygun-Tunç, D., & Tunç, M. N. (2023). A falsificationist treatment of auxiliary hypotheses in social and behavioral sciences: Systematic replications framework. *Meta-Psychology, 7*. https://doi.org/10.15626/MP.2021.2756

Veazie, P. J. (2006). When to combine hypotheses and adjust for multiple tests. *Health Services Research*, *41*(3 pt 1), 804-818. http://doi.org/10.1111/j.1475-6773.2006.00512.x

Venn, J. (1876). *The logic of chance* (2nd ed.). Macmillan and Co.

Wagenmakers, E. J. (2016, September 1). Statistical tools such as p-values and confidence intervals are meaningful only for strictly confirmatory analyses. In turn, preregistration is one. [Comment on the blog post "Why preregistration makes me nervous"]. *Psychological Science.* https://www.psychologicalscience.org/observer/why-preregistration-makes-me-nervous#comment-7860633

Wagenmakers, E. J., Wetzels, R., Borsboom, D., van der Maas, H. L., & Kievit, R. A. (2012). An agenda for purely confirmatory research. *Perspectives on Psychological Science*, *7*(6), 632-638. https://doi.org/10.1177/1745691612463078

Wasserman, L. (2013, March 14). Double misunderstandings about *p*-values. *Normal Deviate.* https://normaldeviate.wordpress.com/2013/03/14/double-misunderstandings-about-p-values/

Wilson, W. (1962). A note on the inconsistency inherent in the necessity to perform multiple comparisons. *Psychological Bulletin*, *59*(4), 296-300. https://doi.org/10.1037/h0040447

Worrall, J. (2010). Theory confirmation and novel evidence. In D. G. Mayo & A. Spanos (Eds.), *Error and inference: Recent exchanges on experimental reasoning, reliability, and the objectivity and rationality of science* (pp. 125-169). Cambridge University Press. http://doi.org/10.1017/CBO9780511657528

---

*Acknowledgements*: I am grateful to members of the academic community on social media for providing feedback on a previous version of this article. I am also grateful to Sander Greenland for his suggestions.

*Funding:* I declare no funding sources.

*Conflict of interest:* I declare no conflict of interest.

*Biography*: I am a professor of psychology at Durham University, UK. In related work, I have argued that it is not always problematic to engage in questionable research practices such as hypothesising after the results are known and uncorrected multiple testing. I have also criticised some science reforms, such as preregistration and stricter adherence to Neyman-Pearson hypothesis testing. For further information about my work in this area, please visit https://sites.google.com/site/markrubinsocialpsychresearch/replication-crisis

*Correspondence:* Correspondence should be addressed to Mark Rubin at the Department of Psychology, Durham University, South Road, Durham, DH1 3LE, UK. E-mail: Mark-Rubin@outlook.com